\documentclass[unpublished,a4paper,onecolumn,superscriptaddress,11pt]{quantumarticle}
\usepackage{tabu}
\usepackage{tikzit}
\usepackage{endnotes}
\tikzstyle{box}=[shape=rectangle, text height=1.5ex, text depth=0.25ex, yshift=0.5mm, fill=white, draw=black, minimum height=5mm, yshift=-0.5mm, minimum width=5mm, font={\small}]
\tikzstyle{Z dot}=[inner sep=0mm, minimum size=2mm, shape=circle, draw=black, fill={rgb,255: red,160; green,255; blue,160}]
\tikzstyle{Z phase dot}=[minimum size=5mm, font={\footnotesize\boldmath}, shape=rectangle, rounded corners=2mm, inner sep=0.2mm, outer sep=-2mm, scale=0.8, tikzit shape=circle, draw=black, fill={rgb,255: red,160; green,255; blue,160}, tikzit draw=blue]
\tikzstyle{X dot}=[Z dot, shape=circle, draw=black, fill={rgb,255: red,220; green,0; blue,0}]
\tikzstyle{X phase dot}=[Z phase dot, tikzit shape=circle, tikzit draw=blue, fill={rgb,255: red,220; green,0; blue,0}, font={\footnotesize\color{white}\boldmath}]
\tikzstyle{hadamard}=[fill=yellow, draw=black, shape=rectangle, inner sep=0.6mm, minimum height=1.5mm, minimum width=1.5mm]
\tikzstyle{vertex}=[inner sep=0mm, minimum size=1mm, shape=circle, draw=black, fill=black]
\tikzstyle{vertex set}=[inner sep=0mm, minimum size=1mm, shape=circle, draw=black, fill=white, font={\footnotesize\boldmath}]

\tikzstyle{hadamard edge}=[-, color=blue, dashed, dash pattern=on 3pt off 1.5pt, thick]
\tikzstyle{brace edge}=[-, tikzit draw=blue, decorate, decoration={brace,amplitude=1mm,raise=-1mm}]
\tikzstyle{diredge}=[->]
\tikzstyle{highlight T}=[-, draw={rgb,255: red,8; green,0; blue,255}, very thick, shorten <=-0.5pt, shorten >=0.5pt]

\input{zx.tikzdefs}
\usepackage{hyperref}
\usepackage{subcaption}
\usepackage{enumitem}
\usepackage{xspace}
\makeatletter
\newcommand\etc{etc\@ifnextchar.{}{.\@}\xspace}
\makeatother

\newcommand{\CZ}{\ensuremath{\textrm{CZ}}\xspace}
\newcommand{\CX}{\ensuremath{\textrm{CNOT}}\xspace}

\newcommand{\CNOT}{\CX}

\usepackage{bm}

\newcommand{\bra}[1]{\ensuremath{\left\langle #1 \right|}}
\newcommand{\ket}[1]{\ensuremath{\left|  #1 \right\rangle}}

\newcommand{\ketbra}[2]{\ensuremath{\ket{#1}\!\bra{#2}}}
\usepackage{amsmath,amsthm,amssymb}
\theoremstyle{definition}
\newtheorem{theorem}{Theorem}[section]

\newtheorem{definition}[theorem]{Definition}
\newtheorem{example}[theorem]{Example}

\newtheorem{algorithm}[theorem]{Algorithm}

\usepackage{rwd-drafting}

\begin{document}

% Author macros::begin %%%%%%%%%%%%%%%%%%%%%%%%%%%%%%%%%%%%%%%%%%%%%%%%

%\title[Quantum Circuit Optimisation by Phase Teleportation]{Quantum Circuit Optimisation by Phase Teleportation}

\title{Reducing T-count with the ZX-calculus}

\author{Aleks Kissinger}
\affiliation{Radboud University Nijmegen}
\email{aleks@cs.ru.nl}
\homepage{https://www.cs.ru.nl/A.Kissinger}

\author{John van de Wetering}
\affiliation{Radboud University Nijmegen}
\email{john@vdwetering.name}
\homepage{http://vdwetering.name}

%\maketitle

\begin{abstract}
Reducing the number of non-Clifford quantum gates present in a circuit is an important task for efficiently implementing quantum computations, especially in the fault-tolerant regime.
We present a new method for reducing the number of T-gates in a quantum circuit based on the ZX-calculus, which matches or beats previous approaches to T-count reduction on the majority of our benchmark circuits in the ancilla-free case, in some cases yielding up to 50\% improvement. Our method begins by representing the quantum circuit as a ZX-diagram, a tensor network-like structure that can be transformed and simplified according to the rules of the ZX-calculus. We then show that a recently-proposed simplification strategy can be extended to reduce T-count using a new technique called phase teleportation. This technique allows non-Clifford phases to combine and cancel by propagating non-locally through a generic quantum circuit. Phase teleportation does not change the number or location of non-phase gates and the method also applies to arbitrary non-Clifford phase gates as well as gates with unknown phase parameters in parametrised circuits. Furthermore, the simplification strategy we use is powerful enough to validate equality of many circuits. In particular, we use it to show that our optimised circuits are indeed equal to the original ones.
We have implemented the routines of this paper in the open-source library PyZX.
\end{abstract}


\section{Introduction}

Quantum circuits give a simple, universal language for describing quantum computations at a low level. It is often useful when studying circuits to distinguish between two kinds of primitive operations: Clifford gates and non-Clifford gates. Circuits consisting only of Clifford gates can be efficiently classically simulated~\cite{aaronsongottesman2004}, and can be implemented in a fault-tolerant manner with relative ease within many quantum error correcting codes such as the surface code~\cite{RaussendorfHarrington,horsman2012surface}. However, achieving universality requires at least one non-Clifford gate, such as the $T$ gate. While techniques such as magic state distillation and injection allow for fault-tolerant implementation of $T$ gates, they typically require an order of magnitude more resources than Clifford gates~\cite{CampbellRoads}. Hence, minimisation of non-Clifford gates within a circuit is of paramount importance to fault-tolerant quantum computation.

There are methods for computing exact-optimal solutions to the problem of T-count minimisation, but they do so at the cost of an exponential running time \cite{amy2013meet,di2016parallelizing}. To date, the most successful scalable approaches to $T$-count minimisation have been based on \textit{phase polynomials}. Such methods rely on an efficient representation of circuits consisting of just CNOT and Z-phase gates in terms of their action on basis states. The first heuristic method for efficiently reducing T-count and T-depth using this representation, called \textit{Tpar}, was introduced by Amy et al in Ref.~\cite{amy2014polynomial}. The results of that paper were later improved upon in Ref.~\cite{amy2016t} and \cite{heyfron2018efficient} by exploiting equivalences between the phase polynomial optimisation problem and other known hard problems: respectively, least-distance Reed-Muller decoding, and least-rank factorisation of certain 3-tensors.

Phase-polynomial methods share the limitation that they cannot deal directly with arbitrary quantum circuits. In particular, an arbitrary circuit will also contain Hadamard gates, which destroy the phase polynomial structure. 
Na\"ively, one can simply cut the circuit into Hadamard-free sections and apply the optimisation locally. This can be significantly improved by preprocessing to produce larger Hadamard-free sections: either by simple gate transformations~\cite{abdessaied2014quantum,nam2018automated} or introducing ancillae and classical control~\cite{heyfron2018efficient}.

While these approaches introduce various tricks and refinements, they share a reliance on phase polynomials as a common core. In this paper, we propose a new approach to reducing non-Clifford gate count based on the theoretical framework laid out in Ref.~\cite{cliff-simp}. We first transform a circuit into a special kind of tensor network called a \emph{ZX-diagram}~\cite{CD1,CD2}. This diagram is then subject to a collection of graphical transformation rules called the \textit{ZX-calculus}~\cite{Backens1}. By breaking the rigid circuit structure, ZX-diagrams are then subject to simplifications that have no circuit analogue.

It was noted in Ref.~\cite{cliff-simp} that non-Clifford phases (i.e. angles which are not multiples of $\pi/2$) form an obstruction to the simplification. To overcome this issue, we introduce one crucial refinement to the simplification procedure: the \textit{gadgetization} of non-Clifford phases. By splitting a node containing a phase into two parts consisting of the node itself, and a new \textit{phase gadget}, phases can propagate non-locally through a ZX-diagram and potentially cancel or combine with each other. In the case where there are no Hadamard gates in the circuit, these gadgets correspond to phase-parity terms in the representation of a phase polynomial, hence this non-local propagation can be seen as a generalisation of phase polynomial techniques to general circuits.

After performing a combination of phase-gadgetization, simplification, and phase-gadget cancellation, we can use a variation on the technique described in Ref.~\cite{cliff-simp} to re-extract a quantum circuit from the ZX-diagram with fewer non-Clifford phases. Alternatively, we can exploit the fact that our simplification procedure is completely parametric in non-Clifford phase angles to do something more lightweight: rather than combining two phase-gadgets into one, we can simply let the angle from one phase gadget `jump' onto the other one: $(\alpha_i, \alpha_j) \leadsto (\alpha_i + \alpha_j, 0)$. Since this doesn't have any effect on the graphical structure of the \zxdiagram, performing this modification to the phases of the original circuit will result in a new circuit that reduces to the same \zxdiagram as before. As a consequence, the new circuit is provably equivalent to the old one.

Hence, rather than re-extracting a circuit from a ZX-diagram, we use it as a tool for discovering phases that can be shifted around non-locally without changing the computed unitary. We call this technique \textit{phase teleportation}. A pleasant property of phase teleportation, as opposed to the simplify-and-extract method, is that it leaves the structure of the quantum circuit completely intact, only changing the parameters. Hence, 2-qubit gate count is never increased and gates are always applied between the same pairs of qubits as before. As pointed out in Ref.~\cite{nam2018automated}, this could be advantageous when the circuit has been designed with limited qubit connectivity of the physical qubits in mind. Both optimisation routines are implemented in the open source Python library \emph{PyZX}~\cite{PyZXGithub}.

By leaving the circuit model we can sometimes `look around' obstructions such as Hadamard gates to find more optimisations. We see this translated in our results. In benchmark circuits with an abundance of Hadamard gates we can significantly outperform previous methods.

We also use the simplification routine for ZX-diagrams to validate equality of circuits. We do this by composing the adjoint of the optimised circuit with the original circuit and checking whether our simplification routine reduces the resulting ZX-diagram to the identity. While this method cannot detect errors in a circuit, the set of rewrite rules forms a certificate of equality when it does reduce a circuit to the identity.
While the general problem of circuit equality validation is QMA-hard~\cite{bookatz2012qmacomplete}, and hence a general efficient validation strategy is unlikely to exist, our method is powerful enough to validate equality of our optimised circuits as well as those produced in Ref.~\cite{nam2018automated}.

It should be noted that we target ancilla-free optimisation and compare ourselves to the best known results for ancilla-free T-count reduction. It is already known that the required amount of T gates can decrease when ancillae are allowed~\cite{amy2014polynomial}. Ref.~\cite{heyfron2018efficient} obtains lower T-counts on many of the circuits we consider by introducing ancillae via a technique called \textit{Hadmard gadgetization}. Using a hybrid of our approach and more advanced phase polynomial techniques, we can obtain ZX-diagrams which \textit{in principle} exhibit very low T-counts, but re-extracting a quantum circuit from the ZX-diagram becomes an obstacle, and will almost certainly require introducing ancillae. This remains an open problem, which we will discuss in more detail in Section~\ref{sec:discussion}.

\medskip

\noindent \textbf{Note}: A few days after the first version of this paper appeared as a preprint~\cite{tcountpreprint}, Zhang and Chen reported nearly identical numbers to those in Table~\ref{fig:results}, using an independent approach based on Pauli rotations~\cite{zhang2019optimizing}. It is a topic of ongoing research as to why these seemingly quite different methods produce the same T-counts.

\medskip

\noindent \textbf{Acknowledgements}: The authors are supported in part by AFOSR grant FA2386-18-1-4028. They would like to thank Earl Campbell and Luke Heyfron for useful discussions and help running the Topt tool as well as Matthew Amy for checking various optimised circuits in the Feynman verifier~\cite{AmyVerification}. They furthermore thank Quanlong Wang and Niel de Beaudrap for interesting discussions on the ZX-calculus, phase gadgets, and circuit optimisation. Finally, they would like to thank Will Zeng and the Unitary Fund for their support of the PyZX project.

\section{Results}

Our procedure simplifies quantum circuits formed from the following primitive set of gates:
\begin{equation}\label{eq:gate-set}
\CNOT \ :=\
\left(\begin{matrix}
  1 & 0 & 0 & 0 \\
  0 & 1 & 0 & 0 \\
  0 & 0 & 0 & 1 \\
  0 & 0 & 1 & 0 \\
\end{matrix}\right)
\qquad
Z_\alpha \ :=\
\left(\begin{matrix}
  1 & 0 \\
  0 & e^{i \alpha}
\end{matrix}\right)
\qquad
H \ :=\ \frac{1}{\sqrt{2}}
\left(\begin{matrix}
  1 & 1 \\
  1 & -1
\end{matrix}\right)
\end{equation}
with the goal of reducing the total number of gates of the form $Z_{\alpha}$ where $\alpha \neq k \cdot \frac\pi2$ for $k \in \mathbb Z$. For all of the benchmark circuits, all of those gates are $T := Z_{\pi/4}$, so from hence forth, we will simply refer to this number as the T-count.

\begin{figure}
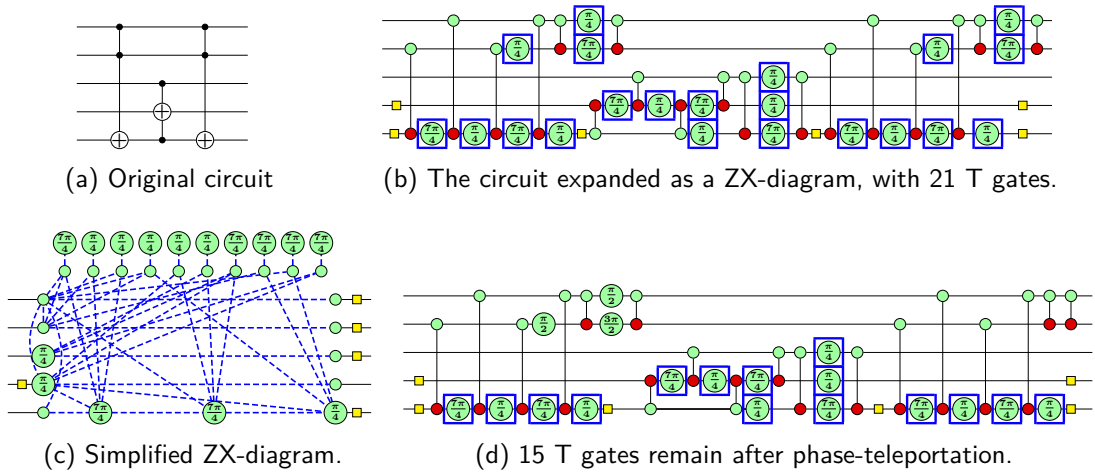
%[!htb]
  \begin{subfigure}[b]{0.35\textwidth}
        \centering
        \scalebox{0.75}{\begin{tikzpicture}
	\begin{pgfonlayer}{nodelayer}
		\node [style=none] (0) at (-3, 2) {};
		\node [style=none] (1) at (-3, 1) {};
		\node [style=none] (2) at (-3, 0) {};
		\node [style=none] (3) at (-3, -1) {};
		\node [style=none] (4) at (-3, -2) {};
		\node [style=none] (5) at (3, 2) {};
		\node [style=none] (6) at (3, 1) {};
		\node [style=none] (7) at (3, 0) {};
		\node [style=none] (8) at (3, -1) {};
		\node [style=none] (9) at (3, -2) {};
		\node [style=vertex set] (12) at (-1.5, -2) {$\!+\!$};
		\node [style=vertex] (13) at (-1.5, 2) {};
		\node [style=vertex] (14) at (-1.5, 1) {};
		\node [style=vertex] (15) at (0, 0) {};
		\node [style=vertex] (16) at (0, -2) {};
		\node [style=vertex set] (17) at (0, -1) {$\!+\!$};
		\node [style=vertex set] (18) at (1.5, -2) {$\!+\!$};
		\node [style=vertex] (19) at (1.5, 2) {};
		\node [style=vertex] (20) at (1.5, 1) {};
		\node [style=none] (21) at (-6.25, 0) {};
		\node [style=none] (22) at (6.5, 0) {};
	\end{pgfonlayer}
	\begin{pgfonlayer}{edgelayer}
		\draw (0.center) to (5.center);
		\draw (1.center) to (6.center);
		\draw (2.center) to (7.center);
		\draw (3.center) to (8.center);
		\draw (4.center) to (9.center);
		\draw (13) to (12);
		\draw (15) to (16);
		\draw (19) to (18);
	\end{pgfonlayer}
\end{tikzpicture}}
        \caption{Original circuit}
        \label{fig:tof3-circuit}
    \end{subfigure}
    \begin{subfigure}[b]{0.6\textwidth}
        \centering
        \scalebox{0.75}{\input{tof3-zx-circuit.tikz}}
        \caption{The circuit expanded as a \zxdiagram, with 21 T gates.}
        \label{fig:tof3-zx-circuit}
    \end{subfigure}
    
    \bigskip

    \begin{subfigure}[b]{0.35\textwidth}
        \centering
        \scalebox{0.75}{\begin{tikzpicture}
	\begin{pgfonlayer}{nodelayer}
		\node [style=none] (0) at (-6.5, 0.25) {};
		\node [style=none] (1) at (-6.5, -0.75) {};
		\node [style=none] (2) at (-6.5, -1.75) {};
		\node [style=none] (3) at (-6.5, -2.75) {};
		\node [style=none] (4) at (-6.5, -3.75) {};
		\node [style=Z dot] (6) at (-5.25, -3.75) {};
		\node [style=Z dot] (7) at (-5.25, -0.75) {};
		\node [style=Z dot] (9) at (-4.5, 1.25) {};
		\node [style=Z dot] (10) at (-5.25, 0.25) {};
		\node [style=Z dot] (12) at (-3.5, 1.25) {};
		\node [style=Z phase dot] (14) at (-3.25, -3.75) {$\frac{7\pi}{4}$};
		\node [style=Z dot] (15) at (-2.5, 1.25) {};
		\node [style=Z dot] (20) at (-1.5, 1.25) {};
		\node [style=Z phase dot] (31) at (0.75, -3.75) {$\frac{7\pi}{4}$};
		\node [style=Z dot] (32) at (-0.5, 1.25) {};
		\node [style=Z phase dot] (33) at (-5.25, -1.75) {$\frac{\pi}{4}$};
		\node [style=Z dot] (35) at (0.5, 1.25) {};
		\node [style=Z dot] (38) at (1.5, 1.25) {};
		\node [style=Z phase dot] (41) at (-5.25, -2.75) {$\frac{\pi}{4}$};
		\node [style=Z dot] (43) at (4.5, 1.25) {};
		\node [style=Z dot] (55) at (2.5, 1.25) {};
		\node [style=Z dot] (61) at (3.5, 1.25) {};
		\node [style=Z phase dot] (64) at (5, -3.75) {$\frac{\pi}{4}$};
		\node [style=none] (74) at (6.25, 0.25) {};
		\node [style=none] (75) at (6.25, -0.75) {};
		\node [style=none] (76) at (6.25, -1.75) {};
		\node [style=none] (77) at (6.25, -2.75) {};
		\node [style=none] (78) at (6.25, -3.75) {};
		\node [style=Z phase dot] (79) at (2.5, 2.25) {$\frac{7\pi}{4}$};
		\node [style=Z phase dot] (80) at (-4.5, 2.25) {$\frac{7\pi}{4}$};
		\node [style=Z phase dot] (81) at (0.5, 2.25) {$\frac{\pi}{4}$};
		\node [style=Z phase dot] (83) at (-2.5, 2.25) {$\frac{\pi}{4}$};
		\node [style=Z phase dot] (84) at (3.5, 2.25) {$\frac{7\pi}{4}$};
		\node [style=Z phase dot] (86) at (-0.5, 2.25) {$\frac{\pi}{4}$};
		\node [style=Z phase dot] (88) at (-3.5, 2.25) {$\frac{\pi}{4}$};
		\node [style=Z phase dot] (89) at (4.5, 2.25) {$\frac{7\pi}{4}$};
		\node [style=Z phase dot] (90) at (-1.5, 2.25) {$\frac{\pi}{4}$};
		\node [style=Z phase dot] (91) at (1.5, 2.25) {$\frac{7\pi}{4}$};
		\node [style=Z dot] (92) at (5, 0.25) {};
		\node [style=Z dot] (93) at (5, -0.75) {};
		\node [style=Z dot] (94) at (5, -1.75) {};
		\node [style=Z dot] (95) at (5, -2.75) {};
		\node [style=hadamard] (96) at (-6, -2.75) {};
		\node [style=hadamard] (97) at (5.75, -3.75) {};
		\node [style=hadamard] (98) at (5.75, 0.25) {};
		\node [style=hadamard] (99) at (5.75, -0.75) {};
		\node [style=hadamard] (100) at (5.75, -1.75) {};
	\end{pgfonlayer}
	\begin{pgfonlayer}{edgelayer}
		\draw (0.center) to (10);
		\draw (1.center) to (7);
		\draw (2.center) to (33);
		\draw (3.center) to (41);
		\draw (4.center) to (6);
		\draw [style=hadamard edge, bend right] (6) to (10);
		\draw [style=hadamard edge] (6) to (14);
		\draw [style=hadamard edge] (7) to (9);
		\draw [style=hadamard edge] (7) to (10);
		\draw [style=hadamard edge] (7) to (12);
		\draw [style=hadamard edge] (7) to (20);
		\draw [style=hadamard edge] (7) to (55);
		\draw [style=hadamard edge] (7) to (93);
		\draw [style=hadamard edge] (9) to (80);
		\draw [style=hadamard edge] (9) to (10);
		\draw [style=hadamard edge] (9) to (14);
		\draw [style=hadamard edge] (10) to (15);
		\draw [style=hadamard edge] (10) to (20);
		\draw [style=hadamard edge, bend right] (10) to (41);
		\draw [style=hadamard edge] (10) to (92);
		\draw [style=hadamard edge] (10) to (61);
		\draw [style=hadamard edge] (10) to (31);
		\draw [style=hadamard edge] (12) to (88);
		\draw [style=hadamard edge] (12) to (14);
		\draw [style=hadamard edge] (14) to (41);
		\draw [style=hadamard edge] (14) to (31);
		\draw [style=hadamard edge] (14) to (15);
		\draw [style=hadamard edge] (15) to (83);
		\draw [style=hadamard edge] (20) to (64);
		\draw [style=hadamard edge] (20) to (90);
		\draw [style=hadamard edge] (31) to (32);
		\draw [style=hadamard edge] (31) to (64);
		\draw [style=hadamard edge] (31) to (35);
		\draw [style=hadamard edge] (31) to (38);
		\draw [style=hadamard edge] (32) to (86);
		\draw [style=hadamard edge] (32) to (41);
		\draw [style=hadamard edge] (33) to (35);
		\draw [style=hadamard edge] (33) to (38);
		\draw [style=hadamard edge] (33) to (43);
		\draw [style=hadamard edge] (33) to (94);
		\draw [style=hadamard edge] (35) to (81);
		\draw [style=hadamard edge] (38) to (41);
		\draw [style=hadamard edge] (38) to (91);
		\draw [style=hadamard edge] (41) to (64);
		\draw [style=hadamard edge] (41) to (43);
		\draw [style=hadamard edge] (41) to (95);
		\draw [style=hadamard edge] (43) to (89);
		\draw [style=hadamard edge] (55) to (64);
		\draw [style=hadamard edge] (55) to (79);
		\draw [style=hadamard edge] (61) to (64);
		\draw [style=hadamard edge] (61) to (84);
		\draw (64) to (78.center);
		\draw (74.center) to (92);
		\draw (75.center) to (93);
		\draw (76.center) to (94);
		\draw (77.center) to (95);
	\end{pgfonlayer}
\end{tikzpicture}} \\
        \caption{Simplified \zxdiagram.}
        \label{fig:tof3-zx-opt}
    \end{subfigure}
    \begin{subfigure}[b]{0.6\textwidth}
        \centering
        \scalebox{0.75}{\input{tof3-circuit-opt.tikz}}
        \caption{15 T gates remain after phase-teleportation.}
        \label{fig:tof3-circuit-opt}
    \end{subfigure}
  \centering
  \caption{Overview of the steps in our phase-teleportation scheme. %We start with a circuit, which we then convert to the Clifford+phase gate set, written as a \zxdiagram. This \zxdiagram is then simplified, while keeping track of which phases fuse, resulting in a circuit with less phase gates.
  }\label{fig:overview}
\end{figure}

\newcommand{\better}{\rowfont{\bf}}
\newcommand{\worse}{\rowfont{\it}}
\newcommand{\plustoddbetter}{\bf}

\begin{table}%[!htb]
\centering
\begin{tabu}{lrrrrrr}
  Circuit & $n$& T& Best prev. & Method & PyZX & PyZX+TODD \\
  \hline
  \better
  adder$_8$ \cite{AmyGithub} & 24 & 399 & 213 & RM$_m$ & 173 & \plustoddbetter 167 \\
  Adder8 \cite{NRSCMGithub} & 23 & 266 & 56 & NRSCM & 56 & 56 \\
  Adder16 \cite{NRSCMGithub} & 47 & 602 & 120 & NRSCM & 120 & 120 \\
  Adder32 \cite{NRSCMGithub} & 95 & 1274 & 248 & NRSCM & 248 & 248 \\
  Adder64 \cite{NRSCMGithub} & 191 & 2618 & 504 & NRSCM & 504 & 504 \\
  barenco-tof$3$ \cite{NRSCMGithub}  & 5 & 28 & 16 & Tpar & 16 & 16 \\
  barenco-tof$4$ \cite{NRSCMGithub}  & 7 & 56 & 28 & Tpar & 28 & 28 \\
  barenco-tof$5$ \cite{NRSCMGithub}  & 9 & 84 & 40 & Tpar & 40 & 40 \\
  barenco-tof$10$ \cite{NRSCMGithub}  & 19 & 224 & 100 & Tpar & 100 & 100 \\
  tof$_3$ \cite{NRSCMGithub} & 5 & 21 & 15 & Tpar & 15 & 15 \\
  tof$_4$ \cite{NRSCMGithub} & 7 & 35 & 23 & Tpar & 23 & 23 \\
  tof$_5$  \cite{NRSCMGithub}& 9 & 49 & 31 & Tpar & 31 & 31 \\
  tof$_{10}$ \cite{NRSCMGithub} & 19 & 119 & 71 & Tpar & 71 & 71 \\
  \worse
  csla-mux$_3$ \cite{NRSCMGithub}  & 15 & 70 & 58 & RM$_r$ & 62 & \plustoddbetter 45 \\
  \worse
  csum-mux$_9$ \cite{NRSCMGithub}  & 30 & 196 & 76 & RM$_r$ & 84 & \plustoddbetter 72 \\
  \better
  cycle17$_3$ \cite{AmyGithub} & 35 & 4739 & 1944 & RM$_m$ & 1797 & 1797 \\
  \worse
  gf($2^4$)-mult \cite{NRSCMGithub} & 12 & 112 & 56 & TODD & 68 & \plustoddbetter 52 \\
  \worse
  gf($2^5$)-mult \cite{NRSCMGithub} & 15 & 175 & 90 & TODD & 115 & \plustoddbetter 86 \\
  \worse
  gf($2^6$)-mult \cite{NRSCMGithub}  & 18 & 252 & 132 & TODD & 150 & \plustoddbetter 122  \\
  \worse
  gf($2^7$)-mult \cite{NRSCMGithub} & 21 & 343 & 185 & TODD & 217 & \plustoddbetter 173 \\
  \worse
  gf($2^8$)-mult \cite{NRSCMGithub} & 24 & 448 & 216 & TODD & 264 & \plustoddbetter 214 \\
  ham15-low \cite{AmyGithub} & 17 & 161 & 97 & Tpar & 97 & 97 \\
  \better
  ham15-med \cite{AmyGithub} & 17 & 574 & 230 & Tpar & 212 & 212 \\
  ham15-high \cite{AmyGithub} & 20 & 2457 & 1019 & Tpar & 1019 & \plustoddbetter 1013 \\
  hwb$_6$ \cite{AmyGithub} & 7 & 105 & 75 & Tpar & 75 & \plustoddbetter 72 \\
  \better
  hwb$_8$ \cite{AmyGithub} & 12 & 5887 & 3531 & RM$_{m\&r}$ & 3517 & \plustoddbetter 3501 \\
  \worse
  mod-mult-55 \cite{NRSCMGithub} & 9 & 49 & 28 & TODD & 35 & \plustoddbetter 20 \\
  mod-red-21 \cite{NRSCMGithub} & 11 & 119 & 73 & Tpar & 73 & 73 \\
  \better
  mod5$_4$ \cite{NRSCMGithub} & 5 & 28 & 16 & Tpar & 8 & \plustoddbetter 7 \\
  \better
  nth-prime$_6$ \cite{AmyGithub} & 9 & 567 & 400 & RM$_{m\&r}$ & 279 & 279 \\
  \worse
  nth-prime$_8$ \cite{AmyGithub} & 12 & 6671 & 4045 & RM$_{m\&r}$ & 4047 & \plustoddbetter 3958 \\
  qcla-adder$_{10}$ \cite{NRSCMGithub} & 36 & 589 & 162 & Tpar & 162 & \plustoddbetter 158 \\
  \worse
  qcla-com$_7$ \cite{NRSCMGithub} & 24 & 203 & 94 & RM$_m$ & 95 & \plustoddbetter 91 \\
  qcla-mod$_7$ \cite{NRSCMGithub} & 26 & 413 & 235${}^\textrm{a}$ & NRSCM & 237 & \plustoddbetter 216 \\
  rc-adder$_6$ \cite{NRSCMGithub} & 14 & 77 & 47 & RM$_{m\&r}$ & 47 & 47 \\
  vbe-adder$_3$ \cite{NRSCMGithub} & 10 & 70 & 24 & Tpar & 24 & 24
  \end{tabu}
  \caption{Benchmark circuits. The columns \emph{$n$} and \emph{T} contain the amount of qubits and T gates in the original circuit. \emph{Best prev.} is the previous best-known ancilla-free T-count for that circuit and \emph{Method} specifies which method was used: \emph{RM$_m$} and \emph{RM$_r$} refer to the \emph{maximum} and \emph{recursive} Reed-Muller decoder of Ref.~\cite{amy2016t}, \emph{Tpar} is Ref.~\cite{amy2014polynomial}, \emph{TODD} is Ref.~\cite{heyfron2018efficient} and \emph{NRSCM} refers to Ref.~\cite{nam2018automated}. \emph{PyZX} and \emph{PyZX + TODD} specify the T-counts produced by the procedures outlined in the Methods section. Numbers shown in bold are better than previous best, and italics are worse. The superscript (a) indicates an error in a previously reported T-count. The error was found using Amy's Feynman tool~\cite{AmyVerification}.
  \label{fig:results}}
\end{table}

The four main steps of our procedure are depicted in Fig.~\ref{fig:overview} and described in detail in Section~\ref{sec:simplifymain}. If a circuit has gates which are not in the gate set~\eqref{eq:gate-set} (as in Fig.~\ref{fig:tof3-circuit}), we first expand those gates in terms of \CNOT, H, and T gates and translate that circuit into a ZX-diagram (Fig.~\ref{fig:tof3-zx-circuit}). We then apply the simplification procedure described in Section~\ref{sec:simplify-zx-full} to obtain a reduced ZX-diagram (Fig.~\ref{fig:tof3-zx-opt}). Finally, we use the data about corresponding phases obtained from this simplification to perform \textit{phase teleportation} in the original circuit to reduce T-count (Fig.~\ref{fig:tof3-circuit-opt}).

For our benchmarks, we have used all of the Clifford+T benchmark circuits from Refs.~\cite{amy2014polynomial,nam2018automated}, minus some of some of the larger members of the \texttt{gf($2^n$)-mult} family. These benchmark circuits were also used in Refs.~\cite{amy2016t,heyfron2018efficient} and include components that are likely to be of interest to quantum algorithms, such as adders or Grover oracles. Of these 36 benchmark circuits, we are at or improving upon the state of the art for 26 circuits ($\sim$72\%), and we improve on the state of the art on 6 ($\sim$17\%). If we apply some simple post-processing afterwards and feed the resulting circuit into the TODD phase polynomial optimiser~\cite{heyfron2018efficient}, we improve on the state of the art for 20 circuits ($\sim$56\%). These two methods seem to complement each other well in the ancilla-free regime, obtaining significantly better numbers than either of the two methods alone, and matching or beating all other methods for every circuit tested. These results are given in Table~\ref{fig:results}.

We achieve an identical T-count to the previously best-known result for 20 of the 36 circuits.
This is interesting, since the methods we use are quite different from previous methods. As a result this can be seen as evidence that those T-counts exist in some kind of `local optimum', at least in the ancilla-free case. The circuits where PyZX seems to do considerably better are ones that contain many Hadamard gates.
The fact that PyZX achieves improvements when there are many Hadamard gates is as expected, as most other successful methods employ a dedicated phase-polynomial optimiser~\cite{amy2014polynomial,amy2016t,nam2018automated,heyfron2018efficient} that is hampered by the existence of Hadamard gates. On the other hand, the only circuits where phase polynomial techniques significantly out-perform our methods are in the \texttt{gf($2^n$)-mult} family. After some simple pre-processing, these circuits have almost no Hadamard gates, hence they are very well-suited to phase polynomial techniques.

It should be noted that while the circuits of Table~\ref{fig:results} are all written in the Clifford+T gate set, our optimisation routine is agnostic to the values of the non-Clifford phases. We have also tested our routine on the quantum Fourier transform circuits of Ref.~\cite{nam2018automated} that include more general non-Clifford phases, and in each case found that our non-Clifford gate count exactly matched their results.

The optimisation routines are implemented in the open source Python library \emph{PyZX}~\cite{PyZXGithub}. All circuit optimisations were run on a consumer laptop. Our method took a few seconds to run for most circuits, with some of the bigger ones taking up to a few minutes. We tested the correctness of the optimisation by generating the matrix of the original and the optimised circuit for thousands of small randomly generated circuits and checking equality, in addition to doing the same for the smaller benchmark circuits. 
We can also use the ZX-calculus for verification of equality~\cite{chancellor2016coherent}. For all benchmark circuits, we composed the original circuit with the adjoint of the optimised one, and then ran our simplification routine on this circuit. In every case, the resulting circuit was reduced to the identity. Since the set of rewrites needed to do this is vastly different then the ones used to produce the original optimised circuit, this is strong evidence that the optimised circuit is correct, as it is very unlikely that an error in our rewrites would cancel itself in this way. With the same validation scheme we have also verified correctness of all the optimised benchmark circuits of Ref.~\cite{nam2018automated}, except for \texttt{qcla-mod$_7$}, which was then shown to be incorrect using the Feynman tool~\cite{AmyVerification}.

\section{Discussion}\label{sec:discussion}

We have implemented a novel quantum circuit optimisation routine that uses \zxdiagrams to go beyond the rigid structure of the circuit model. This routine matches or beats the previous best-known T-count for the majority of the benchmark circuits we have tested. We have furthermore shown that combining our routine with the TODD compiler~\cite{heyfron2018efficient} achieves T-counts that are better than either of these methods separately. Finally, our simplification routine is powerful enough to validate the correctness of our optimised circuits.

There are quite a few ways in which our routine can be improved or be made more versatile. 
Our method currently does not affect the amount of CNOT or Hadamard gates in the circuit. This is because we are not actually making use of the simplified \zxdiagram to extract a new circuit, but we are simply using this diagram to track cancellation of phase gates. It is possible to extract a circuit directly from the \zxdiagram produced by our routine, but at this stage such a circuit often contains more gates than we started out with. For future work, an obvious direction then is to improve our circuit extraction methods to produce better circuits directly from the \zxdiagrams.

Our method currently only supports ancilla-free optimisation. It has been shown~\cite{amy2014polynomial,heyfron2018efficient} that allowing additional ancillae can greatly decrease the required T-count. A promising approach to introducing ancillae into our simplification procedure is the following. We can treat the reduced ZX-diagram as a phase polynomial circuit, where every non-input corresponds to introducing an ancilla in the $\ket{+}$ state and every non-output corresponds to projecting (i.e. post-selecting) onto $\bra{+}$. Indeed we can transform it into a circuit of this form using the \SpiderRule rule of the \zxcalculus (cf.~Section~\ref{sec:zxcalculus}):
\ctikzfig{tof3-zx-opt-anc}
The middle part of the right-hand side can be described as a phase polynomial (cf.~Section~\ref{sec:phase-gadgets}), and hence can be further reduced by powerful phase polynomial techniques such as Reed-Muller decoding~\cite{amy2016t} or 3-tensor factorisation~\cite{heyfron2018efficient}. The resulting circuit contains post-selection and cannot be realised deterministically in general. However, in Ref.\cite{cliff-simp}, we showed that, if certain graph-theoretic constraints are respected, these interior vertices (i.e.~ancillae) can always be eliminated. However, phase polynomial techniques typically break those constraints, so it is an interesting open problem to see if deterministic computation can still be recovered, possibly by introducing measurements and classical control.

While the \zxcalculus forms a powerful language for reasoning about low-level gate sets (e.g. Clifford+T), it can only reason about Toffoli and CCZ gates in an indirect manner, by first translating those gates into Clifford+T representations. The \emph{ZH-calculus}~\cite{zh-calculus} in contrast, makes it straightforward to reason about mid-level quantum gates such as Toffoli and CCZ gates. It then stands to reason that we can achieve further optimisations for circuits defined in terms of these mid-level gates (such as adders and classical oracles), by first doing simplifications in the ZH-calculus, then translating the diagram into a Clifford+T gate set, and doing further simplifications in the \zxcalculus.

It was already noted in the introduction that our simplification is completely parametric in non-Clifford phase angles. Indeed we show the correctness of the phase teleportation routine in Section~\ref{sec:teleport} by working on a representation of a circuit where all non-Clifford angles are replaced with free parameters. The procedure itself then proceeds by combining and eliminating some parameters. An immediate consequence is that our simplification procedure generalises from concrete circuits to parametrised circuits, where the analogue of T-count reduction is elimination of redundant free parameters. This could potentially yield significant improvements in the performance of hybrid classical/quantum algorithms relying on parametrised circuits, such as the quantum variational eigensolver~\cite{peruzzo2014variational}.

One final open question concerns the power of our circuit equality validation scheme, using the ZX-calculus simplifier. We have already noted that this scheme seems to be powerful enough to validate any optimisation made by our simplification routine or the one found in Ref.~\cite{nam2018automated}. It is then an interesting question to explore the exact power (and limitations) of this scheme.

% what the exact limits of this validation scheme are. %When exactly is our simplification routine powerful enough to prove equality of two circuits?

\section{Methods}\label{sec:simplifymain}

In this section we will explain our main contributions in depth, namely how to do T-count optimisation using the \zxcalculus. On a high level this proceeds in the following way:

\begin{itemize}
  \item First we translate a quantum circuit into a \zxdiagram. See Section~\ref{sec:zxcalculus}.
  \item We apply the diagrammatic simplification procedure \textbf{ZX-simplify} laid out in Sections~\ref{sec:simplify-zx-cliff}-\ref{sec:simplify-zx-full}.
  \item Finally, by keeping track of certain simplification steps, and how they affect phases in the original circuit, we will decrease the T-count of the circuit by means of \textit{phase teleportation}. See Section~\ref{sec:teleport}.
\end{itemize}
Section~\ref{sec:TODD} explains our how our PyZX-produced circuit is combined with post-processing and the TODD compiler.

\subsection{Background: the \zxcalculus}\label{sec:zxcalculus}

We will provide a brief overview of the \zxcalculus. For an in-depth
reference see Ref.~\cite{CKbook}.

The \zxcalculus is a diagrammatic language similar to the familiar
quantum circuit notation.  A \emph{\zxdiagram} (or simply
\emph{diagram}) consists of \emph{wires} and \emph{spiders}.  Wires
entering the diagram from the left are \emph{inputs}; wires exiting to
the right are \emph{outputs}.  Given two diagrams we can compose them
by joining the outputs of the first to the inputs of the second, or
form their tensor product by simply stacking the two diagrams.

Spiders are linear operations which can have any number of input or output
wires.  There are two varieties: $Z$ spiders depicted as green dots and $X$ spiders depicted as red dots:\footnote{If you are reading this
  document in monochrome or otherwise have difficulty distinguishing green and red, $Z$ spiders will appear lightly-shaded and $X$ darkly-shaded.}
\[
\small
\hfill \begin{tikzpicture}
	\begin{pgfonlayer}{nodelayer}
		\node [style=Z phase dot] (0) at (0, 0) {$\alpha$};
		\node [style=none] (1) at (1.25, 1) {};
		\node [style=none] (2) at (-1.25, 1) {};
		\node [style=none] (3) at (-1.25, -1) {};
		\node [style=none] (4) at (1.25, -1) {};
		\node [style=none] (5) at (1.25, 0.5) {};
		\node [style=none] (6) at (-1.25, 0.5) {};
		\node [style=none, rotate=90] (7) at (-1, -0.25) {...};
		\node [style=none, rotate=90] (8) at (1, -0.25) {...};
	\end{pgfonlayer}
	\begin{pgfonlayer}{edgelayer}
		\draw [in=-141, out=0, looseness=0.75] (3.center) to (0);
		\draw [in=180, out=-39, looseness=0.75] (0) to (4.center);
		\draw [in=180, out=22, looseness=0.75] (0) to (5.center);
		\draw [in=180, out=39, looseness=0.75] (0) to (1.center);
		\draw [in=0, out=158, looseness=0.75] (0) to (6.center);
		\draw [in=141, out=0, looseness=0.75] (2.center) to (0);
	\end{pgfonlayer}
\end{tikzpicture} \ := \ \ketbra{0...0}{0...0} +
e^{i \alpha} \ketbra{1...1}{1...1} \hfill
\qquad
\hfill \begin{tikzpicture}
	\begin{pgfonlayer}{nodelayer}
		\node [style=X phase dot] (0) at (0, 0) {$\alpha$};
		\node [style=none] (1) at (1.25, 1) {};
		\node [style=none] (2) at (-1.25, 1) {};
		\node [style=none] (3) at (-1.25, -1) {};
		\node [style=none] (4) at (1.25, -1) {};
		\node [style=none] (5) at (1.25, 0.5) {};
		\node [style=none] (6) at (-1.25, 0.5) {};
		\node [style=none, rotate=90] (7) at (-1, -0.25) {...};
		\node [style=none, rotate=90] (8) at (1, -0.25) {...};
	\end{pgfonlayer}
	\begin{pgfonlayer}{edgelayer}
		\draw [in=-141, out=0, looseness=0.75] (3.center) to (0);
		\draw [in=180, out=-39, looseness=0.75] (0) to (4.center);
		\draw [in=180, out=22, looseness=0.75] (0) to (5.center);
		\draw [in=180, out=39, looseness=0.75] (0) to (1.center);
		\draw [in=0, out=158, looseness=0.75] (0) to (6.center);
		\draw [in=141, out=0, looseness=0.75] (2.center) to (0);
	\end{pgfonlayer}
\end{tikzpicture} \ := \ \ketbra{+...+}{+...+} +
e^{i \alpha} \ketbra{-...-}{-...-} \hfill
\]
The diagram as a whole corresponds to a linear map built from the
spiders (and permutations) by the usual composition and tensor product
of linear maps.  As a special case, diagrams with no inputs represent
(unnormalised) state preparations.

\begin{example}\label{ex:basic-maps-and-states}
  We can immediately write down some simple state preparations and
  unitaries in the \zxcalculus:
  \[
  \begin{array}{rclcrcl}
  \begin{tikzpicture}
	\begin{pgfonlayer}{nodelayer}
		\node [style=Z dot] (0) at (-0.5, 0) {};
		\node [style=none] (1) at (0.5, 0) {};
	\end{pgfonlayer}
	\begin{pgfonlayer}{edgelayer}
		\draw (0) to (1.center);
	\end{pgfonlayer}
\end{tikzpicture} & = & \ket{0} + \ket{1} \ \propto \ket{+} &
  \qquad\qquad &
  \begin{tikzpicture}
	\begin{pgfonlayer}{nodelayer}
		\node [style=X dot] (0) at (-0.5, 0) {};
		\node [style=none] (1) at (0.5, 0) {};
	\end{pgfonlayer}
	\begin{pgfonlayer}{edgelayer}
		\draw (0) to (1.center);
	\end{pgfonlayer}
\end{tikzpicture} & = & \ket{+} + \ket{-} \ \propto \ket{0} \\
  &\quad& & & \quad \\
  \begin{tikzpicture}
	\begin{pgfonlayer}{nodelayer}
		\node [style=Z phase dot] (0) at (0, 0) {$\alpha$};
		\node [style=none] (1) at (-1.25, 0) {};
		\node [style=none] (2) at (1.25, 0) {};
	\end{pgfonlayer}
	\begin{pgfonlayer}{edgelayer}
		\draw (1.center) to (0);
		\draw (0) to (2.center);
	\end{pgfonlayer}
\end{tikzpicture} & = & \ketbra{0}{0} + e^{i \alpha} \ketbra{1}{1} =
  Z_\alpha &
  & 
  \begin{tikzpicture}
	\begin{pgfonlayer}{nodelayer}
		\node [style=X phase dot] (0) at (0, 0) {$\alpha$};
		\node [style=none] (1) at (-1.25, 0) {};
		\node [style=none] (2) at (1.25, 0) {};
	\end{pgfonlayer}
	\begin{pgfonlayer}{edgelayer}
		\draw (1.center) to (0);
		\draw (0) to (2.center);
	\end{pgfonlayer}
\end{tikzpicture} & = & \ketbra{+}{+} + e^{i \alpha} \ketbra{-}{-} = X_\alpha
  \end{array}
  \]
  In particular we have the Pauli matrices:
  \[
  \hfill
  \begin{tikzpicture}
	\begin{pgfonlayer}{nodelayer}
		\node [style=Z phase dot] (0) at (0, 0) {$\pi$};
		\node [style=none] (1) at (-1.25, 0) {};
		\node [style=none] (2) at (1.25, 0) {};
	\end{pgfonlayer}
	\begin{pgfonlayer}{edgelayer}
		\draw (1.center) to (0);
		\draw (0) to (2.center);
	\end{pgfonlayer}
\end{tikzpicture} \ \ =\ \  Z \qquad\qquad   \begin{tikzpicture}
	\begin{pgfonlayer}{nodelayer}
		\node [style=X phase dot] (0) at (0, 0) {$\pi$};
		\node [style=none] (1) at (-1.25, 0) {};
		\node [style=none] (2) at (1.25, 0) {};
	\end{pgfonlayer}
	\begin{pgfonlayer}{edgelayer}
		\draw (1.center) to (0);
		\draw (0) to (2.center);
	\end{pgfonlayer}
\end{tikzpicture}\ \ =\ \  X \qquad\qquad
  \hfill
  \]
\end{example}
It will be convenient to introduce a symbol -- a yellow square -- for
the Hadamard gate. This is defined (up to a global phase) by the equation:
\begin{equation}\label{eq:Hdef}
\hfill
\begin{tikzpicture}
	\begin{pgfonlayer}{nodelayer}
		\node [style=none] (0) at (-2.75, 0) {};
		\node [style=none] (1) at (-0.75, 0) {};
		\node [style=hadamard] (2) at (-1.75, 0) {};
		\node [style=none] (3) at (0, 0) {$=$};
		\node [style=none] (4) at (0.75, 0) {};
		\node [style=Z phase dot] (5) at (1.5, 0) {$\frac\pi2$};
		\node [style=X phase dot] (6) at (2.5, 0) {$\frac\pi2$};
		\node [style=Z phase dot] (7) at (3.5, 0) {$\frac\pi2$};
		\node [style=none] (8) at (4.25, 0) {};
	\end{pgfonlayer}
	\begin{pgfonlayer}{edgelayer}
		\draw (0.center) to (2);
		\draw (2) to (1.center);
		\draw (4.center) to (8.center);
	\end{pgfonlayer}
\end{tikzpicture}
\hfill
\end{equation}

We will often use an alternative notation to simplify the diagrams,
and replace a Hadamard between two spiders by a blue dashed edge, as
illustrated below.
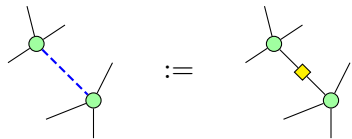
\ctikzfig{blue-edge-def} 
Both the blue edge notation and the Hadamard box can always be
translated back into spiders when necessary. We will refer to the blue
edge as a \emph{Hadamard edge}.

Two diagrams are considered \emph{equal} when one can be deformed to
the other by moving the vertices around in the plane, bending,
unbending, crossing, and uncrossing wires, as long as the connectivity
and the order of the inputs and outputs is maintained. Equivalently, a
ZX-diagram can be considered as a graphical depiction of a tensor network,
as in e.g.~\cite{Penrose}. Then, just like any other tensor network, we can observe that the interpretation of a ZX-diagram is unaffected by deformation. As tensors, Z and X spiders can be written as follows:
\begin{align*}
\left( \  \begin{tikzpicture}
	\begin{pgfonlayer}{nodelayer}
		\node [style=Z phase dot] (0) at (0, 0) {$\alpha$};
	\end{pgfonlayer}
\end{tikzpicture}
 \  \right)_{i_1...i_m}^{j_1...j_n} & =
{\small \begin{cases}
1 & \textrm{ if } i_1 = ... = i_m = j_1 = ... = j_n = 0 \\  
e^{i \alpha} & \textrm{ if } i_1 = ... = i_m = j_1 = ... = j_n = 1 \\
0 & \textrm{ otherwise} 
\end{cases}} \\
\left( \  \begin{tikzpicture}
	\begin{pgfonlayer}{nodelayer}
		\node [style=X phase dot] (0) at (0, 0) {$\alpha$};
	\end{pgfonlayer}
\end{tikzpicture}
 \  \right)_{i_1...i_m}^{j_1...j_n} & =
{\small \frac{1}{\sqrt{2}} \cdot 
\begin{cases}
%1 + (-1)^{{\inlinestyle \sum i_\mu + \sum j_\nu}} \cdot e^{i \alpha}
1 + e^{i \alpha} & \textrm{ if } i_1 \oplus ... \oplus i_m \oplus j_1 \oplus ... \oplus j_n = 0 \\  
1 - e^{i \alpha} & \textrm{ if } i_1 \oplus ... \oplus i_m \oplus j_1 \oplus ... \oplus j_n = 1
\end{cases}}
\end{align*}
where all $i_\alpha, j_\beta$ range over $\{0,1\}$.

\begin{figure}
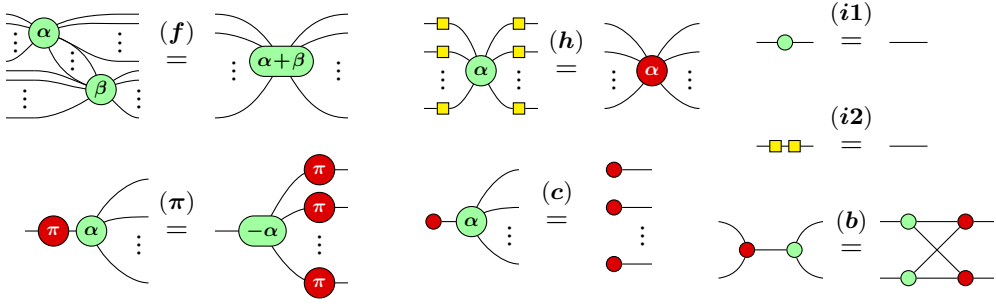

\ctikzfig{ZX-rules}
\vspace{-3mm}
\caption{\label{fig:zx-rules} A convenient presentation for the ZX-calculus. These rules hold
  for all $\alpha, \beta \in [0, 2 \pi)$, and due to $(\bm{h})$ and
  $(\bm{i2})$ all rules also hold with the colours
  interchanged.}
\end{figure}

In addition to simple deformations, ZX-diagrams satisfy a set of equations called the \zxcalculus. There exists several variations of the ZX-calculus. The set of rules we will use is presented in Figure~\ref{fig:zx-rules}. Diagrams that can be transformed into each other using the rules of the ZX-calculus correspond to equal linear maps (up to normalisation). ZX-diagrams with arbitrary angles are expressive enough to represent any linear map~\cite{CD2}. It is often useful to consider \textit{Clifford ZX-diagrams}, by analogy to Clifford circuits, where all angles are restricted to multiples of $\pi/2$. In that case, the rules given in Figure~\ref{fig:zx-rules} are \textit{complete} with respect to linear map equality~\cite{Backens1}. That is, if two Clifford ZX-diagrams describe the same linear map, one can be transformed into the other using the rules in Figure~\ref{fig:zx-rules}. Extensions of the calculus exist that are complete for larger families of \zxdiagrams/maps, including \textit{Clifford+T} \zxdiagrams \cite{SimonCompleteness}, where phases are multiples of $\pi/4$, and arbitrary \zxdiagrams where any phase is allowed~\cite{HarnyAmarCompleteness,JPV-universal,euler-zx}.

 % Equal diagrams correspond to equal linear maps\footnote{We will consider
 %  linear maps $A$ and $B$ to be ``equal'' if they satisfy $A = zB$
 %  for some non-zero $z\in\mathbb{C}$. Backens \cite{Backens:2015aa}
 %  presents a version of the \zxcalculus where equality is ``on the
 %  nose''.}.

% \begin{remark}\label{rem:completeness}
%   The \zxcalculus is \emph{universal} in the sense that any linear map can be expressed as a \zxdiagram. Furthermore, when restricted to \textit{Clifford \zxdiagrams}, i.e. diagrams whose phases are all multiples of $\pi/2$, the version we present in Figure~\ref{fig:zx-rules} is \emph{complete}. That is, for any two Clifford \zxdiagrams that describe the same linear map, there exists a series of rewrites transforming one into the other. Recent extensions to the calculus have been introduced which are complete for the larger 
% \end{remark}

Quantum circuits can be translated into \zxdiagrams in a straightforward manner. 
We will take as our starting point circuits constructed
from the following universal set of gates:
\[
\CNOT \ :=\
\left(\begin{matrix}
  1 & 0 & 0 & 0 \\
  0 & 1 & 0 & 0 \\
  0 & 0 & 0 & 1 \\
  0 & 0 & 1 & 0 \\
\end{matrix}\right)
\qquad\qquad
Z_\alpha \ :=\
\left(\begin{matrix}
  1 & 0 \\
  0 & e^{i \alpha}
\end{matrix}\right)
\qquad\qquad
H \ :=\ \frac{1}{\sqrt{2}}
\left(\begin{matrix}
  1 & 1 \\
  1 & -1
\end{matrix}\right)
\]
This gate set admits a convenient representation in terms of
spiders: 
\begin{align}\label{eq:zx-gates}
\CNOT & = \begin{tikzpicture}
	\begin{pgfonlayer}{nodelayer}
		\node [style=Z dot] (0) at (0, 0.5) {};
		\node [style=none] (1) at (1.25, 0.5) {};
		\node [style=X dot] (2) at (0, -0.5) {};
		\node [style=none] (3) at (-1.25, 0.5) {};
		\node [style=none] (4) at (1.25, -0.5) {};
		\node [style=none] (5) at (-1.25, -0.5) {};
	\end{pgfonlayer}
	\begin{pgfonlayer}{edgelayer}
		\draw (0) to (2);
		\draw (0) to (3.center);
		\draw (0) to (1.center);
		\draw (5.center) to (2);
		\draw (2) to (4.center);
	\end{pgfonlayer}
\end{tikzpicture} &
Z_\alpha & = \begin{tikzpicture}
	\begin{pgfonlayer}{nodelayer}
		\node [style=Z phase dot] (0) at (0, 0) {$\alpha$};
		\node [style=none] (1) at (-1.25, 0) {};
		\node [style=none] (2) at (1.25, 0) {};
	\end{pgfonlayer}
	\begin{pgfonlayer}{edgelayer}
		\draw (1.center) to (0);
		\draw (0) to (2.center);
	\end{pgfonlayer}
\end{tikzpicture} &
H & = \begin{tikzpicture}
	\begin{pgfonlayer}{nodelayer}
		\node [style=none] (0) at (-0.25, 0) {};
		\node [style=none] (1) at (1.75, 0) {};
		\node [style=hadamard] (2) at (0.75, 0) {};
	\end{pgfonlayer}
	\begin{pgfonlayer}{edgelayer}
		\draw (0.center) to (2);
		\draw (2) to (1.center);
	\end{pgfonlayer}
\end{tikzpicture}
\end{align}
Note that for the CNOT gate, the green spider is the first (i.e.~control) qubit and the red spider is the second (i.e.~target) qubit.
Other common gates can easily be expressed in terms of these gates. In particular, $S := Z_{\frac\pi2}$, $T := Z_{\frac\pi4}$ and:
\begin{align}\label{eq:zx-derived-gates}
X_\alpha & = \begin{tikzpicture}
	\begin{pgfonlayer}{nodelayer}
		\node [style=X phase dot] (0) at (0, 0) {$\alpha$};
		\node [style=none] (1) at (-1, 0) {};
		\node [style=none] (2) at (1, 0) {};
		\node [style=Z phase dot] (3) at (-4.25, 0) {$\alpha$};
		\node [style=none] (4) at (-5.5, 0) {};
		\node [style=none] (5) at (-3, 0) {};
		\node [style=none] (6) at (-2, 0) {$=$};
		\node [style=hadamard] (7) at (-3.5, 0) {};
		\node [style=hadamard] (8) at (-5, 0) {};
	\end{pgfonlayer}
	\begin{pgfonlayer}{edgelayer}
		\draw (1.center) to (0);
		\draw (0) to (2.center);
		\draw (4.center) to (3);
		\draw (3) to (5.center);
	\end{pgfonlayer}
\end{tikzpicture} &
\CZ & = \begin{tikzpicture}
	\begin{pgfonlayer}{nodelayer}
		\node [style=Z dot] (0) at (-2.75, 0.5) {};
		\node [style=none] (1) at (-1.5, 0.5) {};
		\node [style=X dot] (2) at (-2.75, -0.5) {};
		\node [style=none] (3) at (-4, 0.5) {};
		\node [style=none] (4) at (-1.5, -0.5) {};
		\node [style=none] (5) at (-4, -0.5) {};
		\node [style=hadamard] (6) at (-3.5, -0.5) {};
		\node [style=hadamard] (7) at (-2, -0.5) {};
		\node [style=none] (8) at (-0.75, 0) {$=$};
		\node [style=Z dot] (9) at (1, 0.5) {};
		\node [style=none] (10) at (2, 0.5) {};
		\node [style=Z dot] (11) at (1, -0.5) {};
		\node [style=none] (12) at (0, 0.5) {};
		\node [style=none] (13) at (2, -0.5) {};
		\node [style=none] (14) at (0, -0.5) {};
		\node [style=hadamard] (15) at (1, 0) {};
		\node [style=none] (16) at (2.75, 0) {$=$};
		\node [style=Z dot] (17) at (4.5, 0.5) {};
		\node [style=none] (18) at (5.5, 0.5) {};
		\node [style=Z dot] (19) at (4.5, -0.5) {};
		\node [style=none] (20) at (3.5, 0.5) {};
		\node [style=none] (21) at (5.5, -0.5) {};
		\node [style=none] (22) at (3.5, -0.5) {};
	\end{pgfonlayer}
	\begin{pgfonlayer}{edgelayer}
		\draw (0) to (2);
		\draw (0) to (3.center);
		\draw (0) to (1.center);
		\draw (5.center) to (2);
		\draw (2) to (4.center);
		\draw (9) to (11);
		\draw (9) to (12.center);
		\draw (9) to (10.center);
		\draw (14.center) to (11);
		\draw (11) to (13.center);
		\draw [style=hadamard edge] (17) to (19);
		\draw (17) to (20.center);
		\draw (17) to (18.center);
		\draw (22.center) to (19);
		\draw (19) to (21.center);
	\end{pgfonlayer}
\end{tikzpicture}
\end{align}

\begin{figure}%[!htb]
  \centering
  \begin{tikzpicture}
	\begin{pgfonlayer}{nodelayer}
		\node [style=Z dot] (0) at (-11.5, 1.5) {};
		\node [style=Z phase dot] (2) at (-10.25, 1.5) {$\alpha$};
		\node [style=Z dot] (3) at (-9, 1.5) {};
		\node [style=X dot] (4) at (-11.5, 0) {};
		\node [style=X dot] (5) at (-9, 0) {};
		\node [style=Z phase dot] (8) at (-7.75, 0) {$\gamma$};
		\node [style=none] (10) at (-13.5, 1.5) {};
		\node [style=none] (11) at (-13.5, 0) {};
		\node [style=none] (12) at (-6.75, 0) {};
		\node [style=none] (13) at (-6.75, 1.5) {};
		\node [style=none] (14) at (-6.75, -1.5) {};
		\node [style=none] (15) at (-13.5, -1.5) {};
		\node [style=hadamard] (173) at (-11.5, -1.5) {};
		\node [style=none] (174) at (-5.75, 0) {$=$};
		\node [style=Z dot] (191) at (-10.25, 0) {};
		\node [style=X dot] (192) at (-10.25, -1.5) {};
		\node [style=Z dot] (196) at (-3.75, 0) {};
		\node [style=Z dot] (197) at (-0.75, 0) {};
		\node [style=Z phase dot] (198) at (0.75, 0) {$\gamma$};
		\node [style=none] (199) at (-4.75, 1.5) {};
		\node [style=none] (200) at (-4.75, 0) {};
		\node [style=none] (201) at (1.5, 0) {};
		\node [style=none] (202) at (1.5, 1.5) {};
		\node [style=none] (203) at (1.5, -1.5) {};
		\node [style=none] (204) at (-4.75, -1.5) {};
		\node [style=Z dot] (207) at (-2.25, 0) {};
		\node [style=Z dot] (208) at (-2.25, -1.5) {};
		\node [style=Z phase dot] (194) at (-2.25, 1.5) {$\alpha$};
		\node [style=hadamard] (215) at (-7.75, 1.5) {};
		\node [style=Z dot] (217) at (0.75, 1.5) {};
		\node [style=hadamard] (218) at (-12.5, 0) {};
		\node [style=Z dot] (235) at (-0.75, -1.5) {};
	\end{pgfonlayer}
	\begin{pgfonlayer}{edgelayer}
		\draw (0) to (4);
		\draw (2) to (3);
		\draw (3) to (5);
		\draw (5) to (8);
		\draw (10.center) to (0);
		\draw (11.center) to (4);
		\draw (4) to (5);
		\draw (8) to (12.center);
		\draw (0) to (2);
		\draw (15.center) to (173);
		\draw (173) to (14.center);
		\draw (191) to (192);
		\draw (200.center) to (196);
		\draw (198) to (201.center);
		\draw [loop] (194) to ();
		\draw (199.center) to (194);
		\draw [loop] (194) to ();
		\draw (3) to (13.center);
		\draw (204.center) to (208);
		\draw (235) to (203.center);
		\draw [style=hadamard edge] (208) to (207);
		\draw [style=hadamard edge] (208) to (235);
		\draw [style=hadamard edge] (207) to (197);
		\draw [style=hadamard edge] (196) to (207);
		\draw [style=hadamard edge] (196) to (194);
		\draw [style=hadamard edge] (197) to (194);
		\draw [style=hadamard edge] (197) to (198);
		\draw [style=hadamard edge] (194) to (217);
		\draw (217) to (202.center);
	\end{pgfonlayer}
\end{tikzpicture}
  \caption{\label{fig:graph-like} A \zxdiagram that comes from a circuit, and its equivalent graph-like \zxdiagram.}
\end{figure}
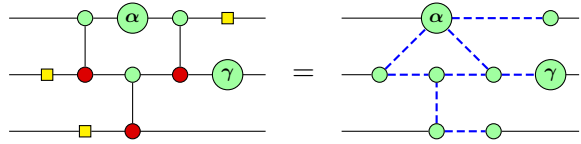

The first step of our simplification procedure is to transform the circuit into something we call a \emph{graph-like} \zxdiagram (see Fig.~\ref{fig:graph-like} for an example).

\begin{definition}\label{def:graph-form}
  A \zxdiagram is \emph{graph-like} when:
  \begin{enumerate}
    \item All spiders are Z-spiders.
    \item Z-spiders are only connected via Hadamard edges.
    \item There are no parallel Hadamard edges or self-loops.
    \item Every input or output is connected to a Z-spider and every Z-spider is connected to at most one input or output.
  \end{enumerate}
\end{definition}

In Ref.~\cite{cliff-simp} it is shown that any \zxdiagram can efficiently be transformed into a graph-like \zxdiagram using the rules of the ZX-calculus. This transformation essentially amounts to turning all X spiders into Z spiders with the \HadamardRule rule, fusing as many spiders together as possible with \SpiderRule, and removing parallel edges/self-loops with the following derived rules:
\begin{equation}\label{eq:parallel-edges-loops}
\begin{tikzpicture}
	\begin{pgfonlayer}{nodelayer}
		\node [style=Z phase dot] (0) at (-1.25, 0) {$\alpha$};
		\node [style=Z phase dot] (1) at (1.25, 0) {$\beta$};
		\node [style=none] (2) at (-2.5, 1) {};
		\node [style=none] (3) at (-2.5, -1) {};
		\node [style=none] (4) at (2.5, 1) {};
		\node [style=none] (5) at (2.5, -1) {};
		\node [style=none] (6) at (-2.5, 0) {...};
		\node [style=none] (7) at (2.5, 0) {...};
		\node [style=none] (10) at (3.5, 0) {=};
		\node [style=Z phase dot] (11) at (5.75, 0) {$\alpha$};
		\node [style=Z phase dot] (12) at (7.25, 0) {$\beta$};
		\node [style=none] (13) at (4.5, 1) {};
		\node [style=none] (14) at (4.5, -1) {};
		\node [style=none] (15) at (8.5, 1) {};
		\node [style=none] (16) at (8.5, -1) {};
		\node [style=none] (17) at (4.5, 0) {...};
		\node [style=none] (18) at (8.5, 0) {...};
	\end{pgfonlayer}
	\begin{pgfonlayer}{edgelayer}
		\draw (0) to (2.center);
		\draw (0) to (3.center);
		\draw (1) to (4.center);
		\draw (1) to (5.center);
		\draw (11) to (13.center);
		\draw (11) to (14.center);
		\draw (12) to (15.center);
		\draw (12) to (16.center);
		\draw [style=hadamard edge, bend right] (0) to (1);
		\draw [style=hadamard edge, bend left] (0) to (1);
	\end{pgfonlayer}
\end{tikzpicture} \qquad
\begin{tikzpicture}
	\begin{pgfonlayer}{nodelayer}
		\node [style=Z phase dot] (0) at (-1.75, 0) {$\alpha$};
		\node [style=none] (1) at (-2.75, -0.75) {};
		\node [style=none] (2) at (-0.75, -0.75) {};
		\node [style=none] (3) at (-1.75, 1) {};
		\node [style=none] (4) at (-1.75, -0.75) {...};
		\node [style=none] (5) at (0, 0) {=};
		\node [style=none] (6) at (1.5, -0.75) {...};
		\node [style=none] (7) at (0.5, -0.75) {};
		\node [style=none] (8) at (2.5, -0.75) {};
		\node [style=Z phase dot] (9) at (1.5, 0) {$\alpha$};
	\end{pgfonlayer}
	\begin{pgfonlayer}{edgelayer}
		\draw (1.center) to (0);
		\draw (0) to (2.center);
		\draw [in=360, out=15, looseness=1.50] (0) to (3.center);
		\draw [in=540, out=165, looseness=1.50] (0) to (3.center);
		\draw (7.center) to (9);
		\draw (9) to (8.center);
	\end{pgfonlayer}
\end{tikzpicture} \qquad
\begin{tikzpicture}
	\begin{pgfonlayer}{nodelayer}
		\node [style=Z phase dot] (0) at (-1.75, 0) {$\alpha$};
		\node [style=none] (1) at (-2.75, -0.75) {};
		\node [style=none] (2) at (-0.75, -0.75) {};
		\node [style=none] (3) at (-1.75, 1) {};
		\node [style=none] (4) at (-1.75, -0.75) {...};
		\node [style=none] (5) at (0, 0) {=};
		\node [style=none] (6) at (1.75, -0.75) {...};
		\node [style=none] (7) at (0.75, -0.75) {};
		\node [style=none] (8) at (2.75, -0.75) {};
		\node [style=Z phase dot] (9) at (1.75, 0) {$\ \alpha + \pi\ $};
	\end{pgfonlayer}
	\begin{pgfonlayer}{edgelayer}
		\draw (1.center) to (0);
		\draw (0) to (2.center);
		\draw [style=hadamard edge, in=360, out=15, looseness=1.50] (0) to (3.center);
		\draw [style=hadamard edge, in=540, out=165, looseness=1.50] (0) to (3.center);
		\draw (7.center) to (9);
		\draw (9) to (8.center);
	\end{pgfonlayer}
\end{tikzpicture}
\end{equation}
In particular, the number of non-Clifford phases in a diagram is never increased, and can actually be decreased by the \SpiderRule rule, as phases are added together.
We call this \textit{graph-like} because the resulting \zxdiagram is essentially an indirected, simple graph whose vertices are labelled by phase angles.

%This transformation involves many spider-fusion steps that already incorporate several commutation identities that are used in other methods, and hence this transformation can often already decrease the amount of phases present in the diagram.

\subsection{Clifford simplification of ZX-diagrams}\label{sec:simplify-zx-cliff}

A spider connected to an input or an output is called a \textit{boundary spider}, whereas all other spiders are called \textit{interior spiders}. If we interpret an $N$-qubit circuit as a ZX-diagram, there are precisely $N$ inputs and $N$ outputs, hence there are at most $2N$ boundary spiders. On the other hand, there will in general be a very large number of interior spiders.

The main idea behind the first part of our simplification strategy is to remove as many interior \textit{Clifford spiders}, i.e. spiders whose phase is a multiple of $\pi/2$, as possible. We do this by using two rewrite rules based on the graph-theoretic operations of \emph{local complementation} and \emph{pivoting}. For the proof of correctness of these rewrite rules we refer to Ref.~\cite{cliff-simp}.

The first rule, based on local complementation, deletes a spider with a phase of $\pm\pi/2$ and introduces edges between all of its neighbours:
\begin{equation}\label{eq:lc-simp}
    \begin{tikzpicture}
	\begin{pgfonlayer}{nodelayer}
		\node [style=Z phase dot] (0) at (-5, 0) {$\ \pm\frac\pi2\ $};
		\node [style=Z phase dot] (2) at (-7.5, -1.25) {$\alpha_1$};
		\node [style=Z phase dot] (4) at (-2.25, -1.25) {$\alpha_n$};
		\node [style=none] (5) at (-8.5, -2.25) {};
		\node [style=none] (6) at (-7.5, -2.25) {};
		\node [style=none] (7) at (-2.25, -2.25) {};
		\node [style=none] (8) at (-1.25, -2.25) {};
		\node [style=none] (9) at (-5, -2) {...};
		\node [style=none] (10) at (-8, -2.25) {...};
		\node [style=none] (11) at (-1.75, -2.25) {...};
		\node [style=none] (16) at (0, -1.5) {$=$};
		\node [style=none] (17) at (10, -1.25) {};
		\node [style=none] (18) at (1.75, -1.25) {};
		\node [style=none] (19) at (9.5, -1.25) {...};
		\node [style=none] (20) at (9, -1.25) {};
		\node [style=Z phase dot] (24) at (2.75, -0.25) {$\ \alpha_1\!\mp\!\frac\pi2\ $};
		\node [style=none] (26) at (2.75, -1.25) {};
		\node [style=none] (27) at (2.25, -1.25) {...};
		\node [style=Z phase dot] (30) at (9, -0.25) {$\ \alpha_n \!\mp\! \frac\pi2\ $};
		\node [style=none] (31) at (-6.5, -4.25) {};
		\node [style=Z phase dot] (32) at (-6.5, -3.25) {$\alpha_2$};
		\node [style=none] (33) at (-7, -4.25) {...};
		\node [style=none] (34) at (-7.5, -4.25) {};
		\node [style=Z phase dot] (35) at (-3.25, -3.25) {$\ \alpha_{n\!-\!1}\ $};
		\node [style=none] (36) at (-3.25, -4.25) {};
		\node [style=none] (37) at (-2.25, -4.25) {};
		\node [style=none] (38) at (-2.75, -4.25) {...};
		\node [style=Z phase dot] (39) at (4.25, -2.75) {$\ \alpha_2\!\mp\!\frac\pi2\ $};
		\node [style=none] (40) at (3.75, -3.75) {...};
		\node [style=none] (41) at (3.25, -3.75) {};
		\node [style=none] (42) at (4.25, -3.75) {};
		\node [style=Z phase dot] (43) at (7.25, -2.75) {$\ \alpha_{n\!-\!1} \!\mp\! \frac\pi2\ $};
		\node [style=none] (44) at (8.25, -3.75) {};
		\node [style=none] (45) at (7.75, -3.75) {...};
		\node [style=none] (46) at (7.25, -3.75) {};
		\node [style=none] (47) at (5.75, -1) {...};
		\node [style=none] (48) at (0, -0.5) {\LcompSimp};
	\end{pgfonlayer}
	\begin{pgfonlayer}{edgelayer}
		\draw (5.center) to (2);
		\draw (6.center) to (2);
		\draw (7.center) to (4);
		\draw (8.center) to (4);
		\draw [style=hadamard edge] (2) to (0);
		\draw [style=hadamard edge] (4) to (0);
		\draw (18.center) to (24);
		\draw (26.center) to (24);
		\draw (20.center) to (30);
		\draw (17.center) to (30);
		\draw (34.center) to (32);
		\draw (31.center) to (32);
		\draw (36.center) to (35);
		\draw (37.center) to (35);
		\draw (41.center) to (39);
		\draw (42.center) to (39);
		\draw (46.center) to (43);
		\draw (44.center) to (43);
		\draw [style=hadamard edge] (24) to (30);
		\draw [style=hadamard edge] (30) to (43);
		\draw [style=hadamard edge] (43) to (39);
		\draw [style=hadamard edge] (39) to (24);
		\draw [style=hadamard edge] (24) to (43);
		\draw [style=hadamard edge] (39) to (30);
		\draw [style=hadamard edge] (0) to (32);
		\draw [style=hadamard edge] (0) to (35);
	\end{pgfonlayer}
\end{tikzpicture}
\end{equation}
Since parallel edges vanish (cf. equations~\eqref{eq:parallel-edges-loops}), this can be seen as complementing the set of edges connecting the neighbours of the deleted vertex, hence the name.

The second rule deletes pairs of \textit{Pauli spiders}, i.e. spiders whose phase is a multiple of $\pi$. For a pair of connected Pauli spiders $u, v$, we can split the neighbourhood of $\{u,v\}$ into three pieces: $U$ the spiders only connected to $u$, $V$ the spiders only connected to $v$, and $W$, the spiders connected to both. We can then delete the pair of spiders $u, v$ provided we introduce complete bipartite graphs on $(U,W)$, $(V,W)$ and $(U,V)$:
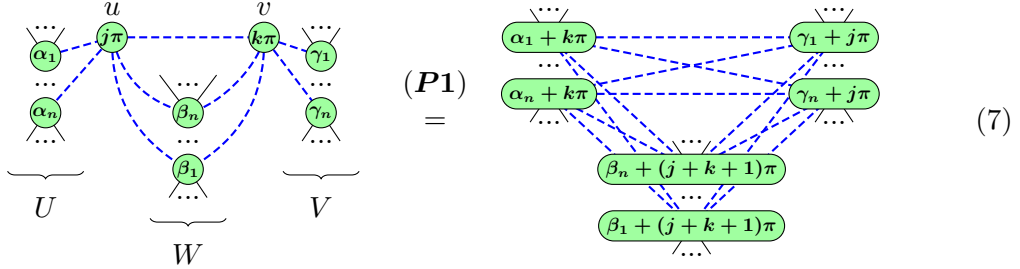
\begin{equation}\label{eq:pivot-simp}
  \begin{tikzpicture}
	\begin{pgfonlayer}{nodelayer}
		\node [style=Z phase dot] (0) at (-8.5, 2.25) {$j\pi$};
		\node [style=Z phase dot] (2) at (-10.25, 1.75) {$\alpha_1$};
		\node [style=none] (16) at (0, 0) {$=$};
		\node [style=Z phase dot] (32) at (-10.25, 0.25) {$\alpha_n$};
		\node [style=Z phase dot] (37) at (-6.5, -1.25) {$\beta_1$};
		\node [style=Z phase dot] (41) at (-6.5, 0.25) {$\beta_n$};
		\node [style=Z phase dot] (45) at (-3, 1.75) {$\gamma_1$};
		\node [style=Z phase dot] (49) at (-3, 0.25) {$\gamma_n$};
		\node [style=Z phase dot] (51) at (-4.5, 2.25) {$k\pi$};
		\node [style=none] (52) at (-10.25, 1) {...};
		\node [style=none] (53) at (-6.5, -0.5) {...};
		\node [style=none] (54) at (-3, 1) {...};
		\node [style=Z phase dot] (55) at (3, 0.75) {$\ \alpha_n+k\pi\ $};
		\node [style=Z phase dot] (56) at (6.75, -1.25) {$\ \beta_n+(j+k+1)\pi\ $};
		\node [style=none] (57) at (6.75, -2) {...};
		\node [style=Z phase dot] (58) at (6.75, -2.75) {$\ \beta_1+(j+k+1)\pi\ $};
		\node [style=Z phase dot] (59) at (10.5, 2.25) {$\ \gamma_1+j\pi\ $};
		\node [style=Z phase dot] (60) at (3, 2.25) {$\ \alpha_1+k\pi\ $};
		\node [style=none] (61) at (10.5, 1.5) {...};
		\node [style=none] (62) at (3, 1.5) {...};
		\node [style=Z phase dot] (63) at (10.5, 0.75) {$\ \gamma_n+j\pi\ $};
		\node [style=none] (64) at (-9.75, 2.5) {};
		\node [style=none] (65) at (-10.75, 2.5) {};
		\node [style=none] (66) at (-10.25, 2.5) {...};
		\node [style=none] (67) at (-10.75, -0.5) {};
		\node [style=none] (68) at (-10.25, -0.5) {...};
		\node [style=none] (69) at (-9.75, -0.5) {};
		\node [style=none] (70) at (-7, -2) {};
		\node [style=none] (71) at (-6.5, -2) {...};
		\node [style=none] (72) at (-6, -2) {};
		\node [style=none] (73) at (-7.25, 1.25) {};
		\node [style=none] (74) at (-6.5, 1) {...};
		\node [style=none] (75) at (-5.75, 1.25) {};
		\node [style=none] (76) at (-3.5, 2.5) {};
		\node [style=none] (77) at (-3, 2.5) {...};
		\node [style=none] (78) at (-2.5, 2.5) {};
		\node [style=none] (79) at (-3, -0.5) {...};
		\node [style=none] (80) at (-3.5, -0.5) {};
		\node [style=none] (81) at (-2.5, -0.5) {};
		\node [style=none] (82) at (2.5, 3) {};
		\node [style=none] (83) at (3, 3) {...};
		\node [style=none] (84) at (3.5, 3) {};
		\node [style=none] (85) at (3, 0) {...};
		\node [style=none] (86) at (2.5, 0) {};
		\node [style=none] (87) at (3.5, 0) {};
		\node [style=none] (88) at (6.25, -0.5) {};
		\node [style=none] (89) at (6.75, -0.5) {...};
		\node [style=none] (90) at (7.25, -0.5) {};
		\node [style=none] (91) at (6.75, -3.5) {...};
		\node [style=none] (92) at (6.25, -3.5) {};
		\node [style=none] (93) at (7.25, -3.5) {};
		\node [style=none] (94) at (10, 3) {};
		\node [style=none] (95) at (10.5, 3) {...};
		\node [style=none] (96) at (11, 3) {};
		\node [style=none] (97) at (10.5, 0) {...};
		\node [style=none] (98) at (10, 0) {};
		\node [style=none] (99) at (11, 0) {};
		\node [style=none] (100) at (0, 1) {\PivotSimp};
		\node [style=none] (101) at (-8.5, 3) {$u$};
		\node [style=none] (102) at (-4.5, 3) {$v$};
		\node [style=none] (103) at (-9.25, -1.5) {};
		\node [style=none] (104) at (-11.25, -1.5) {};
		\node [style=none] (105) at (-10.25, -2.25) {$U$};
		\node [style=none] (106) at (-5.5, -2.75) {};
		\node [style=none] (107) at (-7.5, -2.75) {};
		\node [style=none] (108) at (-6.5, -3.5) {$W$};
		\node [style=none] (109) at (-2, -1.5) {};
		\node [style=none] (110) at (-4, -1.5) {};
		\node [style=none] (111) at (-3, -2.25) {$V$};
	\end{pgfonlayer}
	\begin{pgfonlayer}{edgelayer}
		\draw [style=hadamard edge] (2) to (0);
		\draw [style=hadamard edge] (0) to (32);
		\draw [style=hadamard edge] (0) to (51);
		\draw [style=hadamard edge, bend right] (0) to (37);
		\draw [style=hadamard edge, bend right] (0) to (41);
		\draw [style=hadamard edge, bend left] (51) to (37);
		\draw [style=hadamard edge, bend left, looseness=0.75] (51) to (41);
		\draw [style=hadamard edge] (51) to (45);
		\draw [style=hadamard edge] (51) to (49);
		\draw [style=hadamard edge] (60) to (56);
		\draw [style=hadamard edge] (60) to (58);
		\draw [style=hadamard edge] (55) to (56);
		\draw [style=hadamard edge] (55) to (58);
		\draw [style=hadamard edge] (59) to (56);
		\draw [style=hadamard edge] (59) to (58);
		\draw [style=hadamard edge] (56) to (63);
		\draw [style=hadamard edge] (63) to (58);
		\draw [style=hadamard edge] (60) to (59);
		\draw [style=hadamard edge] (60) to (63);
		\draw [style=hadamard edge] (55) to (63);
		\draw [style=hadamard edge] (55) to (59);
		\draw (64.center) to (2);
		\draw (2) to (65.center);
		\draw (67.center) to (32);
		\draw (69.center) to (32);
		\draw (70.center) to (37);
		\draw (72.center) to (37);
		\draw (41) to (75.center);
		\draw (41) to (73.center);
		\draw (45) to (78.center);
		\draw (45) to (76.center);
		\draw (49) to (80.center);
		\draw (49) to (81.center);
		\draw (60) to (82.center);
		\draw (60) to (84.center);
		\draw (59) to (94.center);
		\draw (59) to (96.center);
		\draw (98.center) to (63);
		\draw (99.center) to (63);
		\draw (86.center) to (55);
		\draw (87.center) to (55);
		\draw (56) to (88.center);
		\draw (56) to (90.center);
		\draw (92.center) to (58);
		\draw (93.center) to (58);
		\draw [style=brace edge] (103.center) to (104.center);
		\draw [style=brace edge] (106.center) to (107.center);
		\draw [style=brace edge] (109.center) to (110.center);
	\end{pgfonlayer}
\end{tikzpicture}
\end{equation}
Again, thanks to \eqref{eq:parallel-edges-loops}, this can be seen as completementing the sets of edges present in the three bipartite graphs $(U,W)$, $(V,W)$ and $(U,V)$.

Since the rules \LcompSimp and \PivotSimp both delete at least one spider, we can simply apply them repeatedly until no rule matches. This gives us a terminating procedure for simplifying our diagram. Note that we do not target the spiders in any specific order. Different orders of application will yield different diagrams (i.e. these rules are not \textit{confluent}), but we always obtain the same amount of non-Clifford spiders at the end.

At this point, the simplification procedure in Ref.~\cite{cliff-simp} employs a variation of \PivotSimp to remove a few more Pauli spiders and terminates. In particular, nothing is done to eliminate \textit{non-Clifford} spiders. This is the goal of the next 2 sections.

\subsection{Phase gadgets}\label{sec:phase-gadgets}

We first introduce a new concept for \zxdiagrams: a \textit{phase gadget}. A phase gadget is simply an arity-1 spider with angle $\alpha$, connected via a Hadamard edge to a spider with no angle:
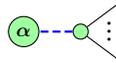
\[
\begin{tikzpicture}
	\begin{pgfonlayer}{nodelayer}
		\node [style=Z dot] (5) at (0.25, 0) {};
		\node [style=none] (6) at (1.25, -0.75) {};
		\node [style=none] (7) at (1.25, 0.75) {};
		\node [style=none, rotate=90] (8) at (1, 0) {...};
		\node [style=Z phase dot] (10) at (-1.25, 0) {$\alpha$};
	\end{pgfonlayer}
	\begin{pgfonlayer}{edgelayer}
		\draw (6.center) to (5);
		\draw (5) to (7.center);
		\draw [style=hadamard edge] (5) to (10);
	\end{pgfonlayer}
\end{tikzpicture}
\]
Phase gadgets are a useful tool for working with \zxdiagrams corresponding to unitaries. For example, the diagram
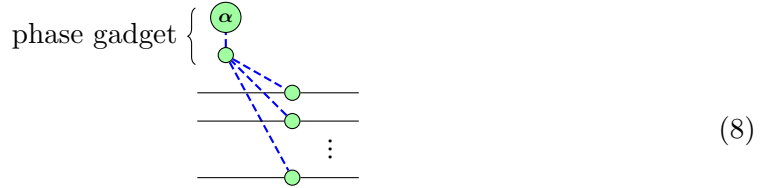
\begin{equation}\label{eq:phase-gadget-unitary}
  \begin{tikzpicture}
	\begin{pgfonlayer}{nodelayer}
		\node [style=Z dot] (5) at (-1.75, 2) {};
		\node [style=Z phase dot] (10) at (-1.75, 3) {$\alpha$};
		\node [style=none] (11) at (-2.5, 1) {};
		\node [style=none] (12) at (-2.5, -1.25) {};
		\node [style=none] (13) at (1.75, 1) {};
		\node [style=none] (14) at (1.75, -1.25) {};
		\node [style=Z dot] (15) at (0, 1) {};
		\node [style=Z dot] (16) at (0, -1.25) {};
		\node [style=none] (41) at (-2.5, 0.25) {};
		\node [style=none] (42) at (1.75, 0.25) {};
		\node [style=Z dot] (45) at (0, 0.25) {};
		\node [style=none, rotate=90] (46) at (1, -0.5) {...};
		\node [style=none] (47) at (-2.75, 3.25) {};
		\node [style=none] (48) at (-2.75, 1.75) {};
		\node [style=none, anchor=east] (49) at (-3, 2.5) {phase gadget};
	\end{pgfonlayer}
	\begin{pgfonlayer}{edgelayer}
		\draw [style=hadamard edge] (5) to (10);
		\draw (11.center) to (15);
		\draw (15) to (13.center);
		\draw (14.center) to (16);
		\draw (16) to (12.center);
		\draw [style=hadamard edge] (5) to (15);
		\draw [style=hadamard edge] (5) to (16);
		\draw (41.center) to (45);
		\draw (45) to (42.center);
		\draw [style=hadamard edge] (45) to (5);
		\draw [style=brace edge] (48.center) to (47.center);
	\end{pgfonlayer}
\end{tikzpicture}
\end{equation}
yields an $n$-qubit unitary $U$ defined by:
\[
U ::
\ket{x_1, ..., x_n} \mapsto
e^{i \alpha (x_1 \oplus \ldots \oplus x_n)} \ket{x_1, ..., x_n}
\]

In fact, it is straightforward to show concretely (or in the ZX-calculus) that this unitary is equal to a ladder of CNOT gates, followed by a single phase gate, followed by the reverse ladder of CNOT gates. For example, on 4 qubits:
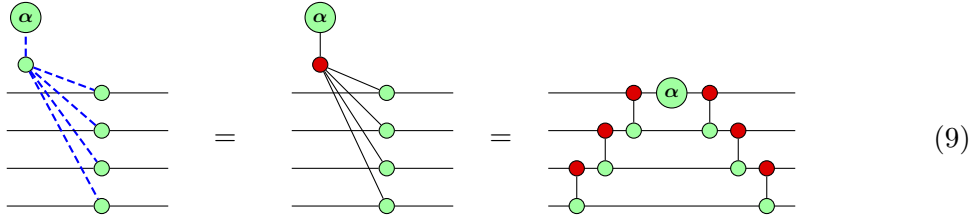
\begin{equation}\label{eq:phasegadget}
\begin{tikzpicture}
	\begin{pgfonlayer}{nodelayer}
		\node [style=Z dot] (5) at (-6, 2) {};
		\node [style=Z phase dot] (10) at (-6, 3.25) {$\alpha$};
		\node [style=none] (11) at (-6.5, 1.25) {};
		\node [style=none] (12) at (-6.5, -0.75) {};
		\node [style=none] (13) at (-2.25, 1.25) {};
		\node [style=none] (14) at (-2.25, -0.75) {};
		\node [style=Z dot] (15) at (-4, 1.25) {};
		\node [style=Z dot] (16) at (-4, -0.75) {};
		\node [style=none] (18) at (-0.75, 0) {=};
		\node [style=Z phase dot] (21) at (1.75, 3.25) {$\alpha$};
		\node [style=none] (22) at (1, 1.25) {};
		\node [style=none] (23) at (1, -0.75) {};
		\node [style=none] (24) at (5.25, 1.25) {};
		\node [style=none] (25) at (5.25, -0.75) {};
		\node [style=Z dot] (26) at (3.5, 1.25) {};
		\node [style=Z dot] (27) at (3.5, -0.75) {};
		\node [style=X dot] (29) at (1.75, 2) {};
		\node [style=none] (30) at (6.5, 0) {=};
		\node [style=none] (41) at (-6.5, 0.25) {};
		\node [style=none] (42) at (-2.25, 0.25) {};
		\node [style=none] (43) at (1, 0.25) {};
		\node [style=none] (44) at (5.25, 0.25) {};
		\node [style=Z dot] (45) at (-4, 0.25) {};
		\node [style=Z dot] (46) at (3.5, 0.25) {};
		\node [style=Z phase dot] (48) at (11, 1.25) {$\alpha$};
		\node [style=none] (49) at (7.75, 1.25) {};
		\node [style=none] (50) at (7.75, -0.75) {};
		\node [style=none] (51) at (14.25, 1.25) {};
		\node [style=none] (52) at (14.25, -0.75) {};
		\node [style=none] (57) at (7.75, 0.25) {};
		\node [style=none] (58) at (14.25, 0.25) {};
		\node [style=Z dot] (60) at (9.25, -0.75) {};
		\node [style=Z dot] (61) at (10, 0.25) {};
		\node [style=X dot] (62) at (9.25, 0.25) {};
		\node [style=X dot] (63) at (10, 1.25) {};
		\node [style=X dot] (64) at (12, 1.25) {};
		\node [style=X dot] (65) at (12.75, 0.25) {};
		\node [style=Z dot] (66) at (12, 0.25) {};
		\node [style=Z dot] (67) at (12.75, -0.75) {};
		\node [style=none] (68) at (-6.5, -1.75) {};
		\node [style=none] (69) at (-2.25, -1.75) {};
		\node [style=Z dot] (70) at (-4, -1.75) {};
		\node [style=none] (71) at (1, -1.75) {};
		\node [style=none] (72) at (5.25, -1.75) {};
		\node [style=Z dot] (73) at (3.5, -1.75) {};
		\node [style=none] (74) at (7.75, -1.75) {};
		\node [style=none] (75) at (14.25, -1.75) {};
		\node [style=Z dot] (78) at (8.5, -1.75) {};
		\node [style=X dot] (79) at (8.5, -0.75) {};
		\node [style=X dot] (80) at (13.5, -0.75) {};
		\node [style=Z dot] (81) at (13.5, -1.75) {};
	\end{pgfonlayer}
	\begin{pgfonlayer}{edgelayer}
		\draw [style=hadamard edge] (5) to (10);
		\draw (11.center) to (15);
		\draw (15) to (13.center);
		\draw (14.center) to (16);
		\draw (16) to (12.center);
		\draw [style=hadamard edge] (5) to (15);
		\draw [style=hadamard edge] (5) to (16);
		\draw (22.center) to (26);
		\draw (26) to (24.center);
		\draw (25.center) to (27);
		\draw (27) to (23.center);
		\draw (29) to (26);
		\draw (29) to (27);
		\draw (29) to (21);
		\draw (41.center) to (45);
		\draw (45) to (42.center);
		\draw (43.center) to (46);
		\draw (46) to (44.center);
		\draw (46) to (29);
		\draw [style=hadamard edge] (45) to (5);
		\draw (49.center) to (51.center);
		\draw (57.center) to (58.center);
		\draw (50.center) to (52.center);
		\draw (62) to (60);
		\draw (61) to (63);
		\draw (64) to (66);
		\draw (65) to (67);
		\draw (69.center) to (70);
		\draw (70) to (68.center);
		\draw [style=hadamard edge] (5) to (70);
		\draw (72.center) to (73);
		\draw (73) to (71.center);
		\draw (29) to (73);
		\draw (74.center) to (75.center);
		\draw (79) to (78);
		\draw (80) to (81);
	\end{pgfonlayer}
\end{tikzpicture}
\end{equation}
Since these gates are diagonal in the computational basis, they commute with each other. This also follows straightforwardly from the \SpiderRule rule:
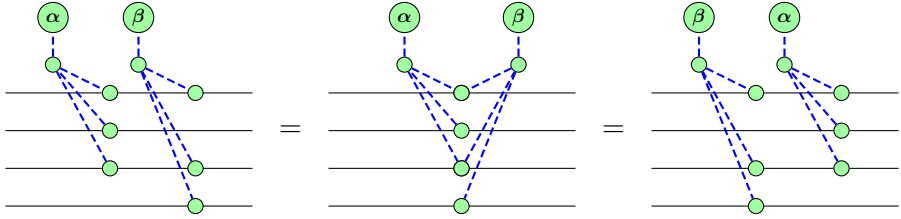
\ctikzfig{phase-gadget-commute}
Arbitrary diagonal unitaries, i.e. unitaries of the form:
\[
U ::
\ket{x_1, ..., x_n} \mapsto
e^{i f(x_1, \ldots, x_n)} \ket{x_1, ..., x_n}
\]
for some $f : \{0,1\}^n \to \mathbb R$, can easily be expressed in terms of phase gadgets. For example:
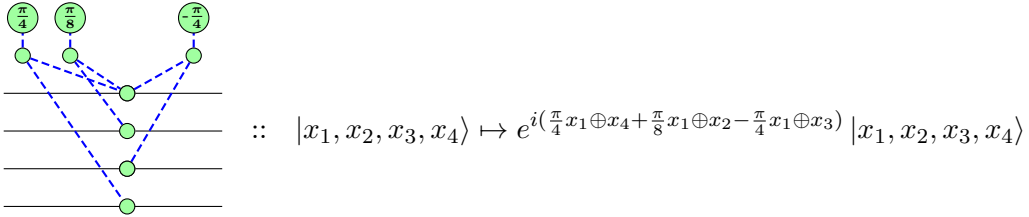
\[
\begin{tikzpicture}
	\begin{pgfonlayer}{nodelayer}
		\node [style=Z dot] (75) at (3, 2) {};
		\node [style=Z phase dot] (76) at (3, 3) {$\frac\pi8$};
		\node [style=none] (77) at (1.25, 1) {};
		\node [style=none] (78) at (1.25, -1) {};
		\node [style=none] (79) at (7, 1) {};
		\node [style=none] (80) at (7, -1) {};
		\node [style=Z dot] (81) at (4.5, 1) {};
		\node [style=Z dot] (82) at (4.5, -1) {};
		\node [style=none] (83) at (1.25, 0) {};
		\node [style=none] (84) at (7, 0) {};
		\node [style=Z dot] (85) at (4.5, 0) {};
		\node [style=none] (86) at (1.25, -2) {};
		\node [style=none] (87) at (7, -2) {};
		\node [style=Z dot] (88) at (1.75, 2) {};
		\node [style=Z phase dot] (89) at (1.75, 3) {$\frac\pi4$};
		\node [style=Z dot] (90) at (4.5, 1) {};
		\node [style=Z dot] (91) at (4.5, -2) {};
		\node [style=Z dot] (92) at (4.5, -1) {};
		\node [style=Z dot] (93) at (6.25, 2) {};
		\node [style=Z phase dot] (94) at (6.25, 3) {-$\frac\pi4$};
	\end{pgfonlayer}
	\begin{pgfonlayer}{edgelayer}
		\draw [style=hadamard edge] (75) to (76);
		\draw (77.center) to (81);
		\draw (81) to (79.center);
		\draw (80.center) to (82);
		\draw (82) to (78.center);
		\draw [style=hadamard edge] (75) to (81);
		\draw (83.center) to (85);
		\draw (85) to (84.center);
		\draw [style=hadamard edge] (85) to (75);
		\draw (86.center) to (87.center);
		\draw [style=hadamard edge] (88) to (89);
		\draw [style=hadamard edge] (88) to (90);
		\draw [style=hadamard edge] (88) to (91);
		\draw [style=hadamard edge] (93) to (94);
		\draw [style=hadamard edge] (90) to (93);
		\draw [style=hadamard edge] (92) to (93);
	\end{pgfonlayer}
\end{tikzpicture}
\ \ ::\ \ 
\ket{x_1,x_2,x_3,x_4}
\mapsto
e^{i (
\frac \pi 4 x_1 \oplus x_4 +
\frac \pi 8 x_1 \oplus x_2 -
\frac \pi 4 x_1 \oplus x_3)}
\ket{x_1,x_2,x_3,x_4}
\]
In fact, the angles appearing in the phase gadgets correspond to the Fourier expansion\footnote{A brief discussion of the form~\eqref{eq:fourier}, and its relation to the usual Fourier transform of a semi-boolean function, can be found in the appendix of Ref.~\cite{amy2018cnot}.} of the semi-boolean function $f$. That is, we can express any function $f : \{0,1\}^n \to \mathbb R$ as follows:
\begin{equation}\label{eq:fourier}
  f(\vec x) = \alpha + \sum_{\vec y} \alpha_{\vec y} (x_1 y_1 \oplus \ldots \oplus x_n y_n)
\end{equation}
where $\vec x, \vec y \in \{ 0, 1 \}^n$ and $\alpha, \alpha_{\vec y} \in \mathbb R$. In the context of diagonal unitaries, $\alpha$ yields a global phase (which we ignore), and each $\alpha_{\vec y}$ corresponds to a phase gadget.

\textit{Phase polynomial} techniques perform transformations on the function $f$ in order to reduce the T-count needed to implement $U$ (or some $U'$ that is Clifford-equivalent to $U$). In the sequel, we will consider not just phase gadgets arising from unitaries such as \eqref{eq:phase-gadget-unitary}, but phase gadgets appearing in arbitary graph-like \zxdiagrams. Hence, our simplification procedure can be seen as a generalisation of phase polynomial techniques.

\subsection{Full simplification of ZX-diagrams}\label{sec:simplify-zx-full}

In this section, we will introduce rules that reduce the number of non-Clifford spiders in the \zxdiagram, and hence the T-count in the resulting circuit.

First, it is worth noting that the \PivotSimp rule from section~\ref{sec:simplify-zx-cliff} was only able to remove an interior Pauli spider adjacent to another interior Pauli spider. We can now introduce two variations of this rule, \PivotSimpGadget and \PivotSimpGadgetBoundary, which together allow us to remove any remaining interior Pauli spider, at the cost of introducing a phase gadget.

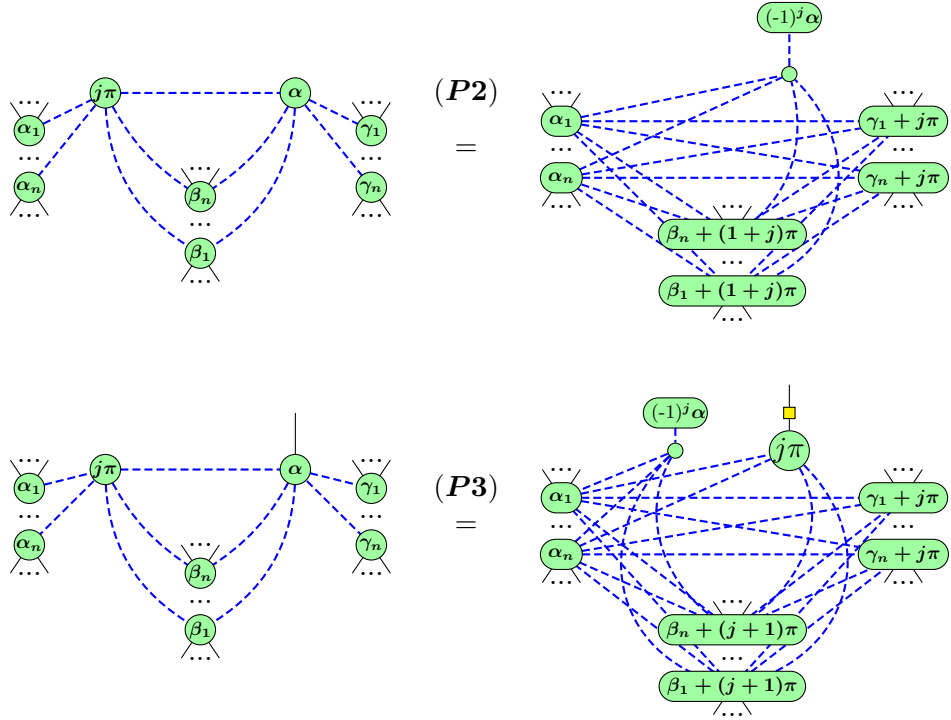
\begin{equation*}\label{eq:pivot-simp-gadget}
\begin{tikzpicture}
	\begin{pgfonlayer}{nodelayer}
		\node [style=Z phase dot] (0) at (-9.5, 1.5) {$j\pi$};
		\node [style=Z phase dot] (2) at (-11.5, 0.5) {$\alpha_1$};
		\node [style=none] (16) at (0, 0) {$=$};
		\node [style=Z phase dot] (32) at (-11.5, -1) {$\alpha_n$};
		\node [style=Z phase dot] (37) at (-7, -2.75) {$\beta_1$};
		\node [style=Z phase dot] (41) at (-7, -1.25) {$\beta_n$};
		\node [style=Z phase dot] (45) at (-2.5, 0.5) {$\gamma_1$};
		\node [style=Z phase dot] (49) at (-2.5, -1) {$\gamma_n$};
		\node [style=Z phase dot] (51) at (-4.5, 1.5) {$\alpha$};
		\node [style=none] (52) at (-11.5, -0.25) {...};
		\node [style=none] (53) at (-7, -2) {...};
		\node [style=none] (54) at (-2.5, -0.25) {...};
		\node [style=Z phase dot] (55) at (2.5, -0.75) {$\ \alpha_n\ $};
		\node [style=Z phase dot] (56) at (7, -2.25) {$\ \beta_n+(1+j)\pi\ $};
		\node [style=none] (57) at (7, -3) {...};
		\node [style=Z phase dot] (58) at (7, -3.75) {$\ \beta_1+(1+j)\pi\ $};
		\node [style=Z phase dot] (59) at (11.5, 0.75) {$\ \gamma_1+j\pi\ $};
		\node [style=Z phase dot] (60) at (2.5, 0.75) {$\ \alpha_1\ $};
		\node [style=none] (61) at (11.5, 0) {...};
		\node [style=none] (62) at (2.5, 0) {...};
		\node [style=Z phase dot] (63) at (11.5, -0.75) {$\ \gamma_n+j\pi\ $};
		\node [style=none] (64) at (-11, 1.25) {};
		\node [style=none] (65) at (-12, 1.25) {};
		\node [style=none] (66) at (-11.5, 1.25) {...};
		\node [style=none] (67) at (-12, -1.75) {};
		\node [style=none] (68) at (-11.5, -1.75) {...};
		\node [style=none] (69) at (-11, -1.75) {};
		\node [style=none] (70) at (-7.5, -3.5) {};
		\node [style=none] (71) at (-7, -3.5) {...};
		\node [style=none] (72) at (-6.5, -3.5) {};
		\node [style=none] (73) at (-7.5, -0.5) {};
		\node [style=none] (74) at (-7, -0.5) {...};
		\node [style=none] (75) at (-6.5, -0.5) {};
		\node [style=none] (76) at (-3, 1.25) {};
		\node [style=none] (77) at (-2.5, 1.25) {...};
		\node [style=none] (78) at (-2, 1.25) {};
		\node [style=none] (79) at (-2.5, -1.75) {...};
		\node [style=none] (80) at (-3, -1.75) {};
		\node [style=none] (81) at (-2, -1.75) {};
		\node [style=none] (82) at (2, 1.5) {};
		\node [style=none] (83) at (2.5, 1.5) {...};
		\node [style=none] (84) at (3, 1.5) {};
		\node [style=none] (85) at (2.5, -1.5) {...};
		\node [style=none] (86) at (2, -1.5) {};
		\node [style=none] (87) at (3, -1.5) {};
		\node [style=none] (88) at (6.5, -1.5) {};
		\node [style=none] (89) at (7, -1.5) {...};
		\node [style=none] (90) at (7.5, -1.5) {};
		\node [style=none] (91) at (7, -4.5) {...};
		\node [style=none] (92) at (6.5, -4.5) {};
		\node [style=none] (93) at (7.5, -4.5) {};
		\node [style=none] (94) at (11, 1.5) {};
		\node [style=none] (95) at (11.5, 1.5) {...};
		\node [style=none] (96) at (12, 1.5) {};
		\node [style=none] (97) at (11.5, -1.5) {...};
		\node [style=none] (98) at (11, -1.5) {};
		\node [style=none] (99) at (12, -1.5) {};
		\node [style=Z phase dot] (101) at (8.5, 3.5) {\ (-1)$^j\alpha$\ };
		\node [style=Z dot] (103) at (8.5, 2) {};
		\node [style=none] (108) at (0, 1.5) {\PivotSimpGadget};
	\end{pgfonlayer}
	\begin{pgfonlayer}{edgelayer}
		\draw [style=hadamard edge] (2) to (0);
		\draw [style=hadamard edge] (0) to (32);
		\draw [style=hadamard edge] (0) to (51);
		\draw [style=hadamard edge, bend right] (0) to (37);
		\draw [style=hadamard edge, bend right=15] (0) to (41);
		\draw [style=hadamard edge, bend left] (51) to (37);
		\draw [style=hadamard edge, bend left=15] (51) to (41);
		\draw [style=hadamard edge] (51) to (45);
		\draw [style=hadamard edge] (51) to (49);
		\draw [style=hadamard edge] (60) to (56);
		\draw [style=hadamard edge] (60) to (58);
		\draw [style=hadamard edge] (55) to (56);
		\draw [style=hadamard edge] (55) to (58);
		\draw [style=hadamard edge] (59) to (56);
		\draw [style=hadamard edge] (59) to (58);
		\draw [style=hadamard edge] (56) to (63);
		\draw [style=hadamard edge] (63) to (58);
		\draw [style=hadamard edge] (60) to (59);
		\draw [style=hadamard edge] (60) to (63);
		\draw [style=hadamard edge] (55) to (63);
		\draw [style=hadamard edge] (55) to (59);
		\draw (64.center) to (2);
		\draw (2) to (65.center);
		\draw (67.center) to (32);
		\draw (69.center) to (32);
		\draw (70.center) to (37);
		\draw (72.center) to (37);
		\draw (41) to (75.center);
		\draw (41) to (73.center);
		\draw (45) to (78.center);
		\draw (45) to (76.center);
		\draw (49) to (80.center);
		\draw (49) to (81.center);
		\draw (60) to (82.center);
		\draw (60) to (84.center);
		\draw (59) to (94.center);
		\draw (59) to (96.center);
		\draw (98.center) to (63);
		\draw (99.center) to (63);
		\draw (86.center) to (55);
		\draw (87.center) to (55);
		\draw (56) to (88.center);
		\draw (56) to (90.center);
		\draw (92.center) to (58);
		\draw (93.center) to (58);
		\draw [style=hadamard edge] (60) to (103);
		\draw [style=hadamard edge] (55) to (103);
		\draw [style=hadamard edge, in=-75, out=30] (56) to (103);
		\draw [style=hadamard edge, in=-45, out=15, looseness=1.25] (58) to (103);
		\draw [style=hadamard edge] (103) to (101);
	\end{pgfonlayer}
\end{tikzpicture}
\end{equation*}

\begin{equation*}\label{eq:pivot-simp-boundary-gadget}
\begin{tikzpicture}
	\begin{pgfonlayer}{nodelayer}
		\node [style=Z phase dot] (0) at (-9.5, 1.5) {$j\pi$};
		\node [style=Z phase dot] (2) at (-11.5, 1) {$\alpha_1$};
		\node [style=none] (16) at (0, 0) {$=$};
		\node [style=Z phase dot] (32) at (-11.5, -0.5) {$\alpha_n$};
		\node [style=Z phase dot] (37) at (-7, -2.75) {$\beta_1$};
		\node [style=Z phase dot] (41) at (-7, -1.25) {$\beta_n$};
		\node [style=Z phase dot] (45) at (-2.5, 1) {$\gamma_1$};
		\node [style=Z phase dot] (49) at (-2.5, -0.5) {$\gamma_n$};
		\node [style=Z phase dot] (51) at (-4.5, 1.5) {$\alpha$};
		\node [style=none] (52) at (-11.5, 0.25) {...};
		\node [style=none] (53) at (-7, -2) {...};
		\node [style=none] (54) at (-2.5, 0.25) {...};
		\node [style=Z phase dot] (55) at (2.5, -0.75) {$\ \alpha_n\ $};
		\node [style=Z phase dot] (56) at (7, -2.75) {$\ \beta_n+(j+1)\pi\ $};
		\node [style=none] (57) at (7, -3.5) {...};
		\node [style=Z phase dot] (58) at (7, -4.25) {$\ \beta_1+(j+1)\pi\ $};
		\node [style=Z phase dot] (59) at (11.5, 0.75) {$\ \gamma_1+j\pi\ $};
		\node [style=Z phase dot] (60) at (2.5, 0.75) {$\ \alpha_1\ $};
		\node [style=none] (61) at (11.5, 0) {...};
		\node [style=none] (62) at (2.5, 0) {...};
		\node [style=Z phase dot] (63) at (11.5, -0.75) {$\ \gamma_n+j\pi\ $};
		\node [style=none] (64) at (-11, 1.75) {};
		\node [style=none] (65) at (-12, 1.75) {};
		\node [style=none] (66) at (-11.5, 1.75) {...};
		\node [style=none] (67) at (-12, -1.25) {};
		\node [style=none] (68) at (-11.5, -1.25) {...};
		\node [style=none] (69) at (-11, -1.25) {};
		\node [style=none] (70) at (-7.5, -3.5) {};
		\node [style=none] (71) at (-7, -3.5) {...};
		\node [style=none] (72) at (-6.5, -3.5) {};
		\node [style=none] (73) at (-7.5, -0.5) {};
		\node [style=none] (74) at (-7, -0.5) {...};
		\node [style=none] (75) at (-6.5, -0.5) {};
		\node [style=none] (76) at (-3, 1.75) {};
		\node [style=none] (77) at (-2.5, 1.75) {...};
		\node [style=none] (78) at (-2, 1.75) {};
		\node [style=none] (79) at (-2.5, -1.25) {...};
		\node [style=none] (80) at (-3, -1.25) {};
		\node [style=none] (81) at (-2, -1.25) {};
		\node [style=none] (82) at (2, 1.5) {};
		\node [style=none] (83) at (2.5, 1.5) {...};
		\node [style=none] (84) at (3, 1.5) {};
		\node [style=none] (85) at (2.5, -1.5) {...};
		\node [style=none] (86) at (2, -1.5) {};
		\node [style=none] (87) at (3, -1.5) {};
		\node [style=none] (88) at (6.5, -2) {};
		\node [style=none] (89) at (7, -2) {...};
		\node [style=none] (90) at (7.5, -2) {};
		\node [style=none] (91) at (7, -5) {...};
		\node [style=none] (92) at (6.5, -5) {};
		\node [style=none] (93) at (7.5, -5) {};
		\node [style=none] (94) at (11, 1.5) {};
		\node [style=none] (95) at (11.5, 1.5) {...};
		\node [style=none] (96) at (12, 1.5) {};
		\node [style=none] (97) at (11.5, -1.5) {...};
		\node [style=none] (98) at (11, -1.5) {};
		\node [style=none] (99) at (12, -1.5) {};
		\node [style=none] (100) at (-4.5, 3) {};
		\node [style=none] (101) at (8.5, 3.75) {};
		\node [style=Z dot] (103) at (8.5, 2) {$j\pi$};
		\node [style=hadamard] (106) at (8.5, 3) {};
		\node [style=Z dot] (107) at (5.5, 2) {};
		\node [style=Z phase dot] (108) at (5.5, 3) {\ (-1)$^j\alpha$\ };
		\node [style=none] (109) at (0, 1) {\PivotSimpGadgetBoundary};
	\end{pgfonlayer}
	\begin{pgfonlayer}{edgelayer}
		\draw [style=hadamard edge] (2) to (0);
		\draw [style=hadamard edge] (0) to (32);
		\draw [style=hadamard edge] (0) to (51);
		\draw [style=hadamard edge, bend right] (0) to (37);
		\draw [style=hadamard edge, bend right=15] (0) to (41);
		\draw [style=hadamard edge, bend left] (51) to (37);
		\draw [style=hadamard edge, bend left=15] (51) to (41);
		\draw [style=hadamard edge] (51) to (45);
		\draw [style=hadamard edge] (51) to (49);
		\draw [style=hadamard edge] (60) to (56);
		\draw [style=hadamard edge] (60) to (58);
		\draw [style=hadamard edge] (55) to (56);
		\draw [style=hadamard edge] (55) to (58);
		\draw [style=hadamard edge] (59) to (56);
		\draw [style=hadamard edge] (59) to (58);
		\draw [style=hadamard edge] (56) to (63);
		\draw [style=hadamard edge] (63) to (58);
		\draw [style=hadamard edge] (60) to (59);
		\draw [style=hadamard edge] (60) to (63);
		\draw [style=hadamard edge] (55) to (63);
		\draw [style=hadamard edge] (55) to (59);
		\draw (64.center) to (2);
		\draw (2) to (65.center);
		\draw (67.center) to (32);
		\draw (69.center) to (32);
		\draw (70.center) to (37);
		\draw (72.center) to (37);
		\draw (41) to (75.center);
		\draw (41) to (73.center);
		\draw (45) to (78.center);
		\draw (45) to (76.center);
		\draw (49) to (80.center);
		\draw (49) to (81.center);
		\draw (60) to (82.center);
		\draw (60) to (84.center);
		\draw (59) to (94.center);
		\draw (59) to (96.center);
		\draw (98.center) to (63);
		\draw (99.center) to (63);
		\draw (86.center) to (55);
		\draw (87.center) to (55);
		\draw (56) to (88.center);
		\draw (56) to (90.center);
		\draw (92.center) to (58);
		\draw (93.center) to (58);
		\draw (51) to (100.center);
		\draw [style=hadamard edge] (60) to (103);
		\draw [style=hadamard edge] (55) to (103);
		\draw [style=hadamard edge, in=30, out=-60] (103) to (56);
		\draw [style=hadamard edge, in=15, out=-45, looseness=1.25] (103) to (58);
		\draw (103) to (106);
		\draw (106) to (101.center);
		\draw [style=hadamard edge] (107) to (60);
		\draw [style=hadamard edge] (107) to (55);
		\draw [style=hadamard edge, in=165, out=-135, looseness=1.25] (107) to (58);
		\draw [style=hadamard edge, in=150, out=-120] (107) to (56);
		\draw [style=hadamard edge] (108) to (107);
	\end{pgfonlayer}
\end{tikzpicture}
\end{equation*}
We apply \PivotSimpGadget when the interior Pauli spider is connected to any other interior spider, while \PivotSimpGadget is applied when it is connected to some boundary spider.
Applying these rules to every remaining interior Pauli spider yields a diagram where every internal spider is either non-Clifford or part of a phase-gadget. If the phase-gadget is Clifford, then it can be removed by either \PivotSimp or by two applications of \LcompSimp. Hence we can reduce to a case where all phase-gadgets are non-Clifford.

We can now apply the following two rules, which both strictly decrease the number of non-Clifford spiders:
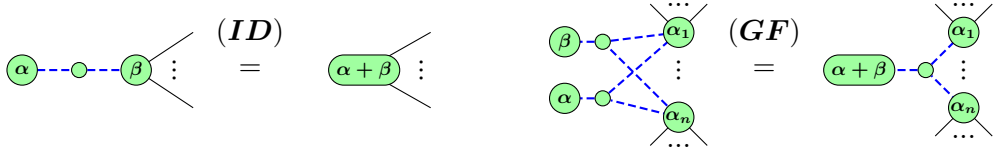
\begin{equation*}
\begin{tikzpicture}
	\begin{pgfonlayer}{nodelayer}
		\node [style=Z phase dot] (0) at (-6, 0) {$\alpha$};
		\node [style=Z phase dot] (1) at (-3, 0) {$\beta$};
		\node [style=Z dot] (2) at (-4.5, 0) {};
		\node [style=none] (3) at (-1.5, 1) {};
		\node [style=none] (4) at (-1.5, -1) {};
		\node [style=none, rotate=90] (8) at (-2, 0) {...};
		\node [style=none] (18) at (0, 0) {=};
		\node [style=Z phase dot] (19) at (3, 0) {$\ \alpha+\beta\ $};
		\node [style=none] (21) at (4.75, 1) {};
		\node [style=none] (22) at (4.75, -1) {};
		\node [style=none, rotate=90] (26) at (4.5, 0) {...};
		\node [style=none] (27) at (0, 1) {\IDSimp};
	\end{pgfonlayer}
	\begin{pgfonlayer}{edgelayer}
		\draw (1) to (3.center);
		\draw (1) to (4.center);
		\draw [style=hadamard edge] (0) to (2);
		\draw [style=hadamard edge] (2) to (1);
		\draw (19) to (21.center);
		\draw (19) to (22.center);
	\end{pgfonlayer}
\end{tikzpicture}\qquad\qquad \begin{tikzpicture}
	\begin{pgfonlayer}{nodelayer}
		\node [style=Z dot] (51) at (-4.25, -0.75) {};
		\node [style=Z phase dot] (52) at (-5.25, -0.75) {$\alpha$};
		\node [style=none] (53) at (-3, 1.75) {};
		\node [style=none] (55) at (-1.5, 1.75) {};
		\node [style=Z dot] (57) at (-2.25, 1) {};
		\node [style=none] (59) at (-3, -2) {};
		\node [style=none] (60) at (-1.5, -2) {};
		\node [style=Z dot] (61) at (-2.25, -1.25) {};
		\node [style=Z dot] (62) at (-4.25, 0.75) {};
		\node [style=Z phase dot] (63) at (-5.25, 0.75) {$\beta$};
		\node [style=Z phase dot] (64) at (-2.25, 1) {$\alpha_1$};
		\node [style=Z phase dot] (66) at (-2.25, -1.25) {$\alpha_n$};
		\node [style=none, rotate=90] (131) at (-2.25, 0) {...};
		\node [style=Z dot] (132) at (4.25, 0) {};
		\node [style=Z phase dot] (133) at (2.5, 0) {$\ \alpha+\beta\ $};
		\node [style=none] (134) at (6, 1.75) {};
		\node [style=none] (135) at (4.5, 1.75) {};
		\node [style=Z phase dot] (136) at (5.25, 1) {$\alpha_1$};
		\node [style=none] (137) at (4.5, -1.75) {};
		\node [style=none] (138) at (6, -1.75) {};
		\node [style=Z dot] (139) at (5.25, -1) {};
		\node [style=Z phase dot] (143) at (5.25, -1) {$\alpha_n$};
		\node [style=none, rotate=90] (144) at (5.25, 0) {...};
		\node [style=none] (145) at (0, 0) {=};
		\node [style=none] (146) at (-2.25, -2) {...};
		\node [style=none] (147) at (-2.25, 1.75) {...};
		\node [style=none] (148) at (5.25, 1.75) {...};
		\node [style=none] (149) at (5.25, -1.75) {...};
		\node [style=none] (150) at (0, 1) {\GadgetSimp};
	\end{pgfonlayer}
	\begin{pgfonlayer}{edgelayer}
		\draw [style=hadamard edge] (51) to (52);
		\draw (53.center) to (57);
		\draw (57) to (55.center);
		\draw [style=hadamard edge] (51) to (57);
		\draw (59.center) to (61);
		\draw (61) to (60.center);
		\draw [style=hadamard edge] (61) to (51);
		\draw [style=hadamard edge] (62) to (63);
		\draw [style=hadamard edge] (62) to (64);
		\draw [style=hadamard edge] (66) to (62);
		\draw [style=hadamard edge] (132) to (133);
		\draw (134.center) to (136);
		\draw (136) to (135.center);
		\draw [style=hadamard edge] (132) to (136);
		\draw (137.center) to (139);
		\draw (139) to (138.center);
		\draw [style=hadamard edge] (139) to (132);
	\end{pgfonlayer}
\end{tikzpicture}
\end{equation*}

When a phase gadget is connected to exactly one other spider, its phase can be combined with the phase on that spider via \IDSimp. This is essentially an application of the rules \IdRule and \HHRule.

When two phase gadgets are connected to exactly the same set of spiders, they can be fused into one via the gadget-fusion rule \GadgetSimp. This rule can be shown using the ZX-calculus:
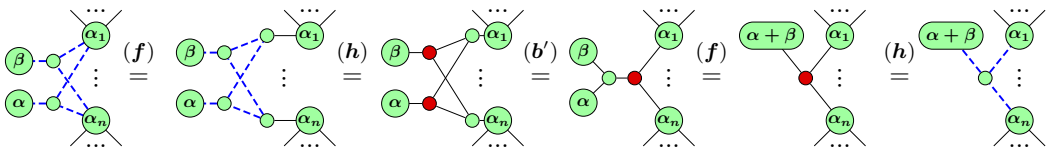
\begin{center}
  \scalebox{0.9}{\begin{tikzpicture}
	\begin{pgfonlayer}{nodelayer}
		\node [style=Z dot] (51) at (-10.5, -0.75) {};
		\node [style=Z phase dot] (52) at (-11.5, -0.75) {$\alpha$};
		\node [style=none] (53) at (-10, 2) {};
		\node [style=none] (55) at (-8.5, 2) {};
		\node [style=none] (59) at (-10, -2) {};
		\node [style=none] (60) at (-8.5, -2) {};
		\node [style=Z dot] (62) at (-10.5, 0.5) {};
		\node [style=Z phase dot] (63) at (-11.5, 0.5) {$\beta$};
		\node [style=Z phase dot] (64) at (-9.25, 1.25) {$\alpha_1$};
		\node [style=Z phase dot] (66) at (-9.25, -1.25) {$\alpha_n$};
		\node [style=none, rotate=90] (131) at (-9.25, 0) {...};
		\node [style=none] (146) at (-9.25, -2) {...};
		\node [style=none] (147) at (-9.25, 2) {...};
		\node [style=none] (148) at (-8, 0) {$=$};
		\node [style=none] (149) at (-8, 0.75) {\spiderrule};
		\node [style=Z dot] (150) at (-5.5, -0.75) {};
		\node [style=Z phase dot] (151) at (-6.5, -0.75) {$\alpha$};
		\node [style=Z dot] (158) at (-5.5, 0.75) {};
		\node [style=Z phase dot] (159) at (-6.5, 0.75) {$\beta$};
		\node [style=Z dot] (160) at (-4.25, 1.25) {};
		\node [style=Z dot] (161) at (-4.25, -1.25) {};
		\node [style=none] (165) at (-3.75, 2) {};
		\node [style=none] (166) at (-2.25, 2) {};
		\node [style=Z dot] (167) at (-3, 1.25) {};
		\node [style=none] (168) at (-3.75, -2) {};
		\node [style=none] (169) at (-2.25, -2) {};
		\node [style=Z dot] (170) at (-3, -1.25) {};
		\node [style=Z phase dot] (171) at (-3, 1.25) {$\alpha_1$};
		\node [style=Z phase dot] (172) at (-3, -1.25) {$\alpha_n$};
		\node [style=none, rotate=90] (173) at (-3.75, 0) {...};
		\node [style=none] (174) at (-3, -2) {...};
		\node [style=none] (175) at (-3, 2) {...};
		\node [style=none] (176) at (-1.75, 0) {$=$};
		\node [style=none] (177) at (-1.75, 0.75) {\hadamardrule};
		\node [style=X dot] (178) at (0.5, -0.75) {};
		\node [style=Z phase dot] (179) at (-0.5, -0.75) {$\alpha$};
		\node [style=X dot] (180) at (0.5, 0.75) {};
		\node [style=Z phase dot] (181) at (-0.5, 0.75) {$\beta$};
		\node [style=Z dot] (182) at (1.75, 1.25) {};
		\node [style=Z dot] (183) at (1.75, -1.25) {};
		\node [style=none] (184) at (1.75, 2) {};
		\node [style=none] (185) at (3.25, 2) {};
		\node [style=Z dot] (186) at (2.5, 1.25) {};
		\node [style=none] (187) at (1.75, -2) {};
		\node [style=none] (188) at (3.25, -2) {};
		\node [style=Z dot] (189) at (2.5, -1.25) {};
		\node [style=Z phase dot] (190) at (2.5, 1.25) {$\alpha_1$};
		\node [style=Z phase dot] (191) at (2.5, -1.25) {$\alpha_n$};
		\node [style=none, rotate=90] (192) at (2, 0) {...};
		\node [style=none] (193) at (2.5, -2) {...};
		\node [style=none] (194) at (2.5, 2) {...};
		\node [style=none] (195) at (3.75, 0) {$=$};
		\node [style=none] (196) at (3.75, 0.75) {\nbialgrule};
		\node [style=Z phase dot] (198) at (5, -0.725) {$\alpha$};
		\node [style=Z phase dot] (200) at (5, 0.775) {$\beta$};
		\node [style=none] (203) at (6.75, 2) {};
		\node [style=none] (204) at (8.25, 2) {};
		\node [style=Z dot] (205) at (7.5, 1.25) {};
		\node [style=none] (206) at (6.75, -2) {};
		\node [style=none] (207) at (8.25, -2) {};
		\node [style=Z dot] (208) at (7.5, -1.25) {};
		\node [style=Z phase dot] (209) at (7.5, 1.25) {$\alpha_1$};
		\node [style=Z phase dot] (210) at (7.5, -1.25) {$\alpha_n$};
		\node [style=none, rotate=90] (211) at (7.5, 0) {...};
		\node [style=none] (212) at (7.5, -2) {...};
		\node [style=none] (213) at (7.5, 2) {...};
		\node [style=X dot] (214) at (6.5, 0) {};
		\node [style=Z dot] (215) at (5.75, 0) {};
		\node [style=none] (216) at (8.75, 0) {$=$};
		\node [style=none] (217) at (8.75, 0.75) {\spiderrule};
		\node [style=none] (220) at (11.75, 2) {};
		\node [style=none] (221) at (13.25, 2) {};
		\node [style=Z dot] (222) at (12.5, 1.25) {};
		\node [style=none] (223) at (11.75, -2) {};
		\node [style=none] (224) at (13.25, -2) {};
		\node [style=Z dot] (225) at (12.5, -1.25) {};
		\node [style=Z phase dot] (226) at (12.5, 1.25) {$\alpha_1$};
		\node [style=Z phase dot] (227) at (12.5, -1.25) {$\alpha_n$};
		\node [style=none, rotate=90] (228) at (12.5, 0) {...};
		\node [style=none] (229) at (12.5, -2) {...};
		\node [style=none] (230) at (12.5, 2) {...};
		\node [style=X dot] (231) at (11.5, 0) {};
		\node [style=Z phase dot] (218) at (10.5, 1.25) {$\ \alpha+\beta\ $};
		\node [style=none] (232) at (14.25, 0) {$=$};
		\node [style=none] (233) at (14.25, 0.75) {\hadamardrule};
		\node [style=none] (234) at (17, 2) {};
		\node [style=none] (235) at (18.5, 2) {};
		\node [style=Z dot] (236) at (17.75, 1.25) {};
		\node [style=none] (237) at (17, -2) {};
		\node [style=none] (238) at (18.5, -2) {};
		\node [style=Z dot] (239) at (17.75, -1.25) {};
		\node [style=Z phase dot] (240) at (17.75, 1.25) {$\alpha_1$};
		\node [style=Z phase dot] (241) at (17.75, -1.25) {$\alpha_n$};
		\node [style=none, rotate=90] (242) at (17.75, 0) {...};
		\node [style=none] (243) at (17.75, -2) {...};
		\node [style=none] (244) at (17.75, 2) {...};
		\node [style=Z dot] (245) at (16.75, 0) {};
		\node [style=Z phase dot] (246) at (15.75, 1.25) {$\ \alpha+\beta\ $};
	\end{pgfonlayer}
	\begin{pgfonlayer}{edgelayer}
		\draw [style=hadamard edge] (51) to (52);
		\draw [style=hadamard edge] (62) to (63);
		\draw [style=hadamard edge] (62) to (64);
		\draw [style=hadamard edge] (66) to (62);
		\draw [style=hadamard edge] (150) to (151);
		\draw [style=hadamard edge] (158) to (159);
		\draw [style=hadamard edge] (158) to (160);
		\draw [style=hadamard edge] (161) to (158);
		\draw (165.center) to (167);
		\draw (167) to (166.center);
		\draw (168.center) to (170);
		\draw (170) to (169.center);
		\draw [style=hadamard edge] (150) to (160);
		\draw [style=hadamard edge] (161) to (150);
		\draw (53.center) to (64);
		\draw (64) to (55.center);
		\draw [style=hadamard edge] (51) to (64);
		\draw (59.center) to (66);
		\draw (66) to (60.center);
		\draw [style=hadamard edge] (66) to (51);
		\draw (160) to (171);
		\draw (161) to (172);
		\draw (178) to (179);
		\draw (180) to (181);
		\draw (180) to (182);
		\draw (183) to (180);
		\draw (184.center) to (186);
		\draw (186) to (185.center);
		\draw (187.center) to (189);
		\draw (189) to (188.center);
		\draw (178) to (182);
		\draw (183) to (178);
		\draw (182) to (190);
		\draw (183) to (191);
		\draw (203.center) to (205);
		\draw (205) to (204.center);
		\draw (206.center) to (208);
		\draw (208) to (207.center);
		\draw (210) to (214);
		\draw (214) to (209);
		\draw (214) to (215);
		\draw (215) to (200);
		\draw (215) to (198);
		\draw (220.center) to (222);
		\draw (222) to (221.center);
		\draw (223.center) to (225);
		\draw (225) to (224.center);
		\draw (227) to (231);
		\draw (231) to (226);
		\draw (231) to (218);
		\draw [loop] (218) to ();
		\draw (234.center) to (236);
		\draw (236) to (235.center);
		\draw (237.center) to (239);
		\draw (239) to (238.center);
		\draw [style=hadamard edge] (241) to (245);
		\draw [style=hadamard edge] (245) to (240);
		\draw [style=hadamard edge] (245) to (246);
		\draw [loop] (246) to ();
	\end{pgfonlayer}
\end{tikzpicture}}
\end{center}
where \NBialgRule is the $n$-ary generalisation of the rule \BialgRule, which follows from the other rules (see e.g.~\cite{CKbook}, \S9.4). For unitaries of the form~\eqref{eq:phase-gadget-unitary}, it corresponds to a well-known simplification used in phase-polynomial circuits, where two phases acting on the same parity of the input qubits can be summed together.

%In the case of our diagrams, it applies whenever two phase-gadgets have exactly the same set of targets.

Each of the rewrite rules \IDSimp and \GadgetSimp removes a non-Clifford spider, and transforms another non-Clifford spider into a Clifford spider, which can be removed by one of the previous rules. We can now fully describe our simplification procedure for graph-like \zxdiagrams.

\begin{algorithm}
    \textbf{ZX-simplify}: Starting with a graph-like ZX-diagram, do the following:
    \begin{enumerate}
        \item Apply \LcompSimp until all interior proper Clifford vertices are removed.
        \item Apply \PivotSimp, \PivotSimpGadget and \PivotSimpGadgetBoundary until all interior Pauli vertices are removed or transformed into phase-gadgets.
        \item Remove all Clifford phase-gadgets using \LcompSimp and \PivotSimp.
        \item Apply \IDSimp and \GadgetSimp wherever possible. If any matches were found, go back to step 1, otherwise we are done.
    \end{enumerate}
\end{algorithm}

This algorithm always terminates as every step either removes a spider or a phase-gadget. 
In terms of complexity we see that if the original diagram had $n$ spiders, that this algorithm takes at most $n$ steps. Each step might need us to toggle the connectivity of all the neighbours of the involved spider. As this spider has at most $n$ neighbours, this could involve $n^2$ operations on the diagram. The complexity of the algorithm is therefore bounded above by $O(n^3)$ elementary graph operations. In practice though, the \zxdiagrams resulting from quantum circuits will be quite sparse, and we tend to see a time scaling roughly between $O(n)$ and $O(n^2)$ on our benchmark circuits.

It will be useful to have a name for the diagrams produced by this simplification procedure.

\begin{definition}
  We say a graph-like \zxdiagram is in \emph{reduced gadget form} when 
  \begin{itemize}
    \item Every internal spider is a non-Clifford spider or part of a non-Clifford phase-gadget.
    \item Every phase-gadget has more than one target.
    \item No two phase-gadgets have the same set of targets.
  \end{itemize}
\end{definition}

\subsection{Phase teleportation}\label{sec:teleport}

The simplification procedure described in the previous section produces a \zxdiagram which does not look at all like a circuit. In order to get a new, simplified circuit out, we could apply (a variation of) the circuit extraction procedure described in Ref.~\cite{cliff-simp}. Alternatively, we can short-circuit the extraction using a trick we refer to as \textit{phase teleportation}.

We begin by replacing every non-Clifford phase in our starting circuit $C$ with a fresh variable name, $\alpha_1, \ldots, \alpha_n$, and storing the angles in a separate table $\tau : \{1, \ldots, n\} \to \mathbb R$.

We can then perform the simplification procedure described in the previous section \textit{symbolically}. That is, we work on a \zxdiagram whose spiders are labelled not just with phase angles, but with polynomials over the variables $(\alpha_1, \ldots, \alpha_n)$.

Then, consider what happens when two variables are added together during the \IDSimp and \GadgetSimp rules. One of two things can occur: $(a)$ the two variables have the same sign or $(b)$ they have different signs:
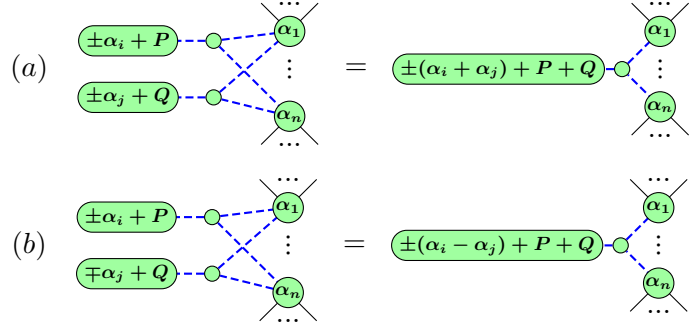
\[
(a) \ \ \ \begin{tikzpicture}
	\begin{pgfonlayer}{nodelayer}
		\node [style=Z dot] (51) at (-4.25, 0.75) {};
		\node [style=Z phase dot] (52) at (-6.5, 0.75) {$\ \pm\alpha_i + P\ $};
		\node [style=none] (53) at (-3, 1.75) {};
		\node [style=none] (55) at (-1.5, 1.75) {};
		\node [style=Z dot] (57) at (-2.25, 1) {};
		\node [style=none] (59) at (-3, -2) {};
		\node [style=none] (60) at (-1.5, -2) {};
		\node [style=Z dot] (61) at (-2.25, -1.25) {};
		\node [style=Z dot] (62) at (-4.25, -0.75) {};
		\node [style=Z phase dot] (63) at (-6.5, -0.75) {$\ \pm\alpha_j + Q\ $};
		\node [style=Z phase dot] (64) at (-2.25, 1) {$\alpha_1$};
		\node [style=Z phase dot] (66) at (-2.25, -1.25) {$\alpha_n$};
		\node [style=none, rotate=90] (131) at (-2.25, 0) {...};
		\node [style=Z dot] (132) at (6.5, 0) {};
		\node [style=Z phase dot] (133) at (3.25, 0) {$\ \pm(\alpha_i+\alpha_j) + P + Q\ $};
		\node [style=none] (134) at (8.25, 1.75) {};
		\node [style=none] (135) at (6.75, 1.75) {};
		\node [style=Z phase dot] (136) at (7.5, 1) {$\alpha_1$};
		\node [style=none] (137) at (6.75, -1.75) {};
		\node [style=none] (138) at (8.25, -1.75) {};
		\node [style=Z dot] (139) at (7.5, -1) {};
		\node [style=Z phase dot] (143) at (7.5, -1) {$\alpha_n$};
		\node [style=none, rotate=90] (144) at (7.5, 0) {...};
		\node [style=none] (145) at (-0.5, 0) {=};
		\node [style=none] (146) at (-2.25, -2) {...};
		\node [style=none] (147) at (-2.25, 1.75) {...};
		\node [style=none] (148) at (7.5, 1.75) {...};
		\node [style=none] (149) at (7.5, -1.75) {...};
	\end{pgfonlayer}
	\begin{pgfonlayer}{edgelayer}
		\draw [style=hadamard edge] (51) to (52);
		\draw (53.center) to (57);
		\draw (57) to (55.center);
		\draw [style=hadamard edge] (51) to (57);
		\draw (59.center) to (61);
		\draw (61) to (60.center);
		\draw [style=hadamard edge] (61) to (51);
		\draw [style=hadamard edge] (62) to (63);
		\draw [style=hadamard edge] (62) to (64);
		\draw [style=hadamard edge] (66) to (62);
		\draw [style=hadamard edge] (132) to (133);
		\draw (134.center) to (136);
		\draw (136) to (135.center);
		\draw [style=hadamard edge] (132) to (136);
		\draw (137.center) to (139);
		\draw (139) to (138.center);
		\draw [style=hadamard edge] (139) to (132);
	\end{pgfonlayer}
\end{tikzpicture}
\]
\[
(b) \ \ \ \begin{tikzpicture}
	\begin{pgfonlayer}{nodelayer}
		\node [style=Z dot] (51) at (-4.25, 0.75) {};
		\node [style=Z phase dot] (52) at (-6.5, 0.75) {$\ \pm\alpha_i + P\ $};
		\node [style=none] (53) at (-3, 1.75) {};
		\node [style=none] (55) at (-1.5, 1.75) {};
		\node [style=Z dot] (57) at (-2.25, 1) {};
		\node [style=none] (59) at (-3, -2) {};
		\node [style=none] (60) at (-1.5, -2) {};
		\node [style=Z dot] (61) at (-2.25, -1.25) {};
		\node [style=Z dot] (62) at (-4.25, -0.75) {};
		\node [style=Z phase dot] (63) at (-6.5, -0.75) {$\ \mp\alpha_j + Q\ $};
		\node [style=Z phase dot] (64) at (-2.25, 1) {$\alpha_1$};
		\node [style=Z phase dot] (66) at (-2.25, -1.25) {$\alpha_n$};
		\node [style=none, rotate=90] (131) at (-2.25, 0) {...};
		\node [style=Z dot] (132) at (6.5, 0) {};
		\node [style=Z phase dot] (133) at (3.25, 0) {$\ \pm(\alpha_i-\alpha_j) + P + Q\ $};
		\node [style=none] (134) at (8.25, 1.75) {};
		\node [style=none] (135) at (6.75, 1.75) {};
		\node [style=Z phase dot] (136) at (7.5, 1) {$\alpha_1$};
		\node [style=none] (137) at (6.75, -1.75) {};
		\node [style=none] (138) at (8.25, -1.75) {};
		\node [style=Z dot] (139) at (7.5, -1) {};
		\node [style=Z phase dot] (143) at (7.5, -1) {$\alpha_n$};
		\node [style=none, rotate=90] (144) at (7.5, 0) {...};
		\node [style=none] (145) at (-0.5, 0) {=};
		\node [style=none] (146) at (-2.25, -2) {...};
		\node [style=none] (147) at (-2.25, 1.75) {...};
		\node [style=none] (148) at (7.5, 1.75) {...};
		\node [style=none] (149) at (7.5, -1.75) {...};
	\end{pgfonlayer}
	\begin{pgfonlayer}{edgelayer}
		\draw [style=hadamard edge] (51) to (52);
		\draw (53.center) to (57);
		\draw (57) to (55.center);
		\draw [style=hadamard edge] (51) to (57);
		\draw (59.center) to (61);
		\draw (61) to (60.center);
		\draw [style=hadamard edge] (61) to (51);
		\draw [style=hadamard edge] (62) to (63);
		\draw [style=hadamard edge] (62) to (64);
		\draw [style=hadamard edge] (66) to (62);
		\draw [style=hadamard edge] (132) to (133);
		\draw (134.center) to (136);
		\draw (136) to (135.center);
		\draw [style=hadamard edge] (132) to (136);
		\draw (137.center) to (139);
		\draw (139) to (138.center);
		\draw [style=hadamard edge] (139) to (132);
	\end{pgfonlayer}
\end{tikzpicture}
\]
Since none of our simplifications will copy any of the variables we started with, these are the only occurences of $\alpha_i$ and $\alpha_j$ in the \zxdiagram. Hence, in case $(a)$, if we replace $\alpha_i$ with $\alpha_i + \alpha_j$ and $\alpha_j$ with $0$, we get an equivalent diagram. %In fact, we can already make this substitution in the starting circuit $C$.

Put another way, in case $(a)$, we can update our table $\tau$ by setting $\tau'(i) := \tau(i) + \tau(j)$, $\tau'(j) := 0$, and $\tau'(k) := \tau(k)$ for $k \notin \{i,j\}$.
Then, $(C, \tau)$ and $(C, \tau')$ describe circuits which are provably equivalent by the rules of \zxcalculus. Case $(b)$ is similar, except we should set $\tau'(i) := \tau(i) - \tau(j)$.

This observation yields the following algorithm:

\begin{algorithm}
  \textbf{Phase teleportation}: Staring with a circuit, do the following:
  \begin{enumerate}
    \item Choose unique variables $\alpha_1, \ldots, \alpha_n$ for each non-Clifford phase and store the pair $(C, \tau)$, where $C$ is the parametrised circuit and $\tau : \{1, \ldots, n\} \to \mathbb R$ assigns each variable to its phase.
    \item Interpret $C$ as a \zxdiagram and run the \textbf{ZX-simplify} algorithm on the simplified diagram while doing the following:
    \begin{quote}
      Whenever \IDSimp or \GadgetSimp are applied to a pair of vertices or phase-gadgets containing variables $\alpha_i$ and $\alpha_j$, respectively, update the phase table $\tau$ as described for cases $(a)$ and $(b)$ above.
    \end{quote}
    \item When \textbf{ZX-simplify} finishes, the pair $(C, \tau')$ describes an equivalent circuit.
  \end{enumerate}
\end{algorithm}

Even though we do compute the reduced gadget form of the circuit $C$, the new circuit we output has the same structure as $C$ itself, but with some of the phases changed. As a result, no new gates are introduced, but many non-Clifford phase gates will have their angles set to $0$ or to multiples of $\pi/2$. Hence, running a dedicated gate minimising circuit optimisation routine afterwards will often be much more effective.

\subsection{Circuit optimisation and TODD}\label{sec:TODD}

We now briefly describe a combined optimisation routing consisting of first running the phase teleportation procedure, then doing some simple post-processing, and finally applying the \emph{TODD} algorithm described in Ref.~\cite{heyfron2018efficient}. %In this section we will describe how these last two steps are applied.

The circuit post-processing works by doing forward and backward passes through the circuit. During the forward pass, we commute 1-qubit gates as far forward as possible using standard gate commutation rules, cancelling and combining gates whenever we can.
% using the rules in Fig.~\ref{fig:commutesimple} and cancel 1- and 2-qubit whenever possible using the rules in Fig.~\ref{fig:rewritesimple}. SWAP gates introduced by the cancellation rules are removed by re-indexing qubits.
We then take the adjoint of the circuit and repeat the process, and keep repeating the process until no more gates are removed.

We then apply the ancilla-free version of the TODD algorithm using the C++ tool Topt~\cite{Topt}. This tool is designed to optimise CNOT+Phase circuits, so we first cut our circuit into Hadamard-free chunks. Then, before running Topt on each chunk, we again use standard gate commutation laws to pull as many gates as possible from neighbouring chunks into the current one. Since Topt is non-deterministic, we run it multiple times and we take the best result. Running Topt on each chunk in this manner then yields the T-counts reported in the last column of Table~\ref{fig:results}.

% wish to make Hadamard-free sub-circuits as large as possible.

% implements advanced heuristics for optimising the T-count of phase-polynomial circuits. We have implemented their algorithm in PyZX. 

% Given a Clifford+T circuit, we first optimise the T-count by our phase-teleportation scheme. Then we optimise the Hadamard count by running the scheme described above. We then wish to make Hadamard-free sub-circuits that are as large as possible, so that TODD has more chances to find optimisations on these sub-circuits. 

% To do so, we again start consuming gates from the left, and we push as many gates as possible past Hadamard gates. In contrast to the previous algorithm, when we cannot commute a gate past a Hadamard, we do not reset the stack, but we simply stop consuming gates on that qubit. Once we can no longer consume gates on any qubit, we are left with a phase-polynomial circuit on the stack on which we run TODD. We run TODD multiple times on the sub-circuit, as the implementation we use is non-deterministic. We take the best result from among these runs.

% After TODD finishes on this sub-circuit, we commute as many gates as possible back to their previous location to the right of the Hadamard gates, so that they can take part in a next round of TODD. Starting with these gates on the stack, we again start consuming until we are stuck, and we again apply TODD on the resulting sub-circuit. This procedure is repeated until the end of the circuit is reached.

\bibliographystyle{plain}
\bibliography{main}

\end{document}